\documentclass[aps,floatfix,showpacs,showkeys,twocolumn]{revtex4}
\usepackage{amssymb}
\usepackage{amsmath}
\usepackage{textcomp}
\usepackage{rcs}
\RCS $Date$
\usepackage{comment}
\usepackage{graphicx}

\newcommand{\be}{\begin{eqnarray}} 
\newcommand{\ee}{\end{eqnarray}} 
\newcommand{\bml}{\begin{multline}}
\newcommand{\eml}{\end{multline}}

\begin{document}

\title{Phonon spectral functions of photo-generated hot carrier plasmas: \\
effects of carrier screening and plasmon--phonon coupling 
}
\author{Kyung-Soo Yi}
\email[Corresponding author. e-mail: ]{ksyi@pusan.ac.kr}
\author{Hye-Jung Kim}
\altaffiliation{Present address: Basic Science Research Institute, University of Ulsan, Ulsan 680-749, Korea}
\affiliation{Department of Physics,
Pusan National University, Busan 46241, Korea}
\date{\RCSDate}

\begin{abstract}
We investigate spectral behavior of phonon spectral functions in an interacting multi-component hot carrier plasma. 
Dielectric polarization functions are formulated so that they satisfy Dyson equations of the effective interactions among plasma components. 
We find a useful sum rule giving simple relation between plasma-species resolved dielectric functions. 
Spectral analysis of various phonon spectral functions is performed considering carrier-phonon channels of polar and nonpolar optical phonons, acoustic deformation-potential, and piezoelectric Coulomb couplings.
Effects of phonon self-energy corrections are examined at finite temperature within a random phase approximation extended to include the effects of dynamic screening, plasmon--phonon coupling, and local-field corrections of the plasma species.
We provide numerical data for the case of a photo-generated electron-hole plasma formed in a wurtzite GaN.
Our result shows clear significance of the multiplicity of the plasma species in the dielectric response and phonon spectral functions of a multi-component plasma.

\end{abstract}
\maketitle

\section{Introduction}

Photo-generated hot electrons and holes in semiconductors form a multi-component plasma (mcp) separated in the reciprocal space and have recently gained renewed interest because of their practical importance in application to short-wavelength optoelectronic devices \cite{Kyhm2011,Hagele,Ozgur,Sarua,Ishioka}.
In solid state plasmas formed in doped polar semiconductors, longitudinal phonons and plasmons are strongly coupled through their macroscopic electric fields, and if the frequency of plasma is comparable to that of the phonon, the mode coupling of these excitations is maximized leading to the formation of coupled plasmon--phonon modes \cite{Quinn-Yi}.
The coupling problems of these excitations between the longitudinal optic (LO) phonons and plasmons in a single-component plasma (scp) such as of doped semiconductors have been investigated extensively both theoretically and experimentally \cite{Abstreiter1984} including experimental confirmation in Raman scattering measurements \cite{Pinczuk1977,Abstreiter1979,Romanek}.  

In a mcp, different plasma species responds differently to an external disturbance resulting in nonidentical multiple polarization functions.
In a multi-component electron--hole plasma, in addition to the well-known compressive optic plasmon--LO phonon coupling of the scp \cite{DasSarma1990}, low-frequency acoustic (AC) collective oscillations also occur \cite{Platzman1973}.
The acoustic plasmon mode is a low-frequency plasma density oscillation due to in-phase motion of electrons and holes similar to the case of AC phonons \cite{Pines,Pinczuk1981,Esperidiao}.
Since the number of plasma components modifies the response functions of the system, the mcp would support a great variety of spectral behavior giving rise to novel features (absent in the scp) such as the coupling of individual plasmons to polar optical phonons.
The mode coupling properties are expected to depend on the plasma species, because the dispersive behavior of self-sustaining oscillations varies for different plasma species \cite{Ishioka}.

Phonon spectral function of a solid state plasma is characterized by the behavior of coupled plasmon--phonon modes through carrier-phonon interaction (in addition to the individual properties of phonons and carriers) and depends strongly on the nature of the screening and collective behavior of the carriers.
For example, 
Jain and co-workers proposed a coupling of LO phonons to quasiparticle excitations (QPE) and studied its effect on the hot electron energy loss rate at low temperature in a single-component electron plasma \cite{Jain1988}.
However, little is known about the nature of the low energy coupled phonon--QPE modes, and the spectral behavior of the QPE modes.
Since there are several polarization functions in the case of a mcp, a careful examination of the spectral behavior is required. 
The spectral behavior in a mcp is expected to be quite involved and different from that of a single-component one \cite{Yi-Kim-CAP2015}. 
The effects of dynamic screening of plasma species on collisional broadening of the phonon spectral function need be clarified in a systematic way in a mcp--phonon coupled system.

\begin{table*}[t]
\caption{Material parameters used for a wurtzite GaN}
\begin{tabular}{p{50pt}p{190pt}p{60pt}p{50pt}} 
\hline\hline
	&	& GaN &	references\\ 
\hline
$\epsilon_\infty$ &optical dielectric constant & 5.35 & \cite{Pau} \\
$\epsilon_0$ &static dielectric constant & 9.7 & \cite{Landolt} \\
$\omega_{\mathrm{LO}}$ & LO phonon energy (meV) & 92 &\cite{Grahn}\\
$\omega_{\mathrm{TO}}$ & TO phonon energy (meV) & 66 &\cite{Grahn}\\
$m_{\rm c}/m_0$ & electron effective mass & 0.22 & \cite{Kyhm} \\
$m_{\mathrm{hh}}/m_0$ & heavy hole effective mass & 1.30 & \cite{Kyhm} \\
$m_{\mathrm{lh}}/m_0$ & light-hole effective mass & 0.3 & \cite{Kyhm} \\
$\rho$ & average mass density (g/cm$^{3}$) & 6.15 &\cite{Landolt} \\
$\mathcal{E}_c$ & conduction band acoustical \par deformation potential (eV) & 5 & \cite{Gil,Tchounkeu,Chuang}\\ 
$\mathcal{E}_v$ & valence band acoustical \par deformation potential (eV) & 2 & \cite{Walle} \\
$\mathcal{D}$ & optical deformation potential (eV/cm) & $10^9$ & \cite{Guess} \\
$\mathit s$ & mean sound velocity (cm/s) & $7.67 \times 10^5$ & \cite{Nakamura} \\
\hline
\end{tabular}
\vspace{3pt}
\end{table*}
 
The purpose of this paper is to map out various phonon spectral functions of a mcp in the $\omega$--$q$ plane and analyze their behavior in detail to elucidate the effects of dynamic screening and various plasmon-phonon coupling at finite temperature. 
The dynamic screening of the carrier--phonon interactions is included by extending the random phase approximation (rpa) to take into account local-field corrections (LFCs) of carriers, and carrier--phonon coupling channels of polar and nonpolar optical phonons are considered. 
We present numerical results applied to the case of electrons and holes optically generated in a solid focusing our study on the influence of dynamic screening and plasmon--phonon coupling on various phonon spectral functions.
A brief account of some preliminary analysis on dielectric responses in a mcp has recently been reported \cite{YI&Kim_PhyscaB}. 
We briefly review the formulation of dielectric responses and phonon spectral function of a mcp in Sec. II and discuss results in Sec. III for the cases of optically excited wurtzite GaN. The main conclusion of our study is in Sec. IV. 
Table I shows the physical parameters of a wurtzite GaN used in our numerical calculation of polarization and dielectric response functions for an idealized mcp.
In GaN, the formation of hot electron-hole plasma is expected for carrier densities larger than $10^{18}-10^{19} \rm cm^{-3}$ \cite{Binet1999}. 
In a wurtzite GaN, the band extrema of conduction and valence bands are located at the center of the Brillouin zone, the former being of $\Gamma_7$ symmetry and the latter splitting into heavy-hole, light-hole, and split-off bands \cite{GaN}.
As one increases the degree of photo excitations, the number of plasma species can be tuned from a two-component (conduction electron--heavy hole) plasma at weak excitation and a three-component (conduction electron--heavy hole--light hole) one at strong excitation.

\section{Formulation}
We consider a mcp produced in an intrinsic semiconductor, by optical excitations, generating electrons in the conduction band and various holes in the valence bands.
  
Here we describe the formulation of dielectric response functions of a mcp and their application to a study of phonon spectral functions of the system. 
The phonon spectral behavior of a solid state plasma would dependent on the carrier screening and scattering channels (distinguished by a parameter $j$) of various carrier--phonon interaction $H_{\rm c-ph}^{(j)}$ via a specific coupling of carrier $(\rm c)$ and phonon (ph). 
Here $j$ denotes various types of phonons with either polar or nonpolar character.
Specific scattering channel dependent carrier--phonon interaction is written as \cite{Bruus,Mahan72,ksyi2007} 
\begin{equation}
H_{\rm c-ph}^{(j)} = \sum_{k q\nu} M_{q,\nu}^{(j)} {\hat n}_{k+q,\nu}\left(a_{q}^{(j)}+ a_{-q}^{(j)\dagger} \right), \label{Heph}
\end{equation}
where $\hat n_{k+q,\nu} (=c_{k+q,\nu}^\dagger c_{k\nu})$ and $M_{q,\nu}^{(j)} ( =M_{-q,\nu}^{(j)*}$) are the Fourier transform of the carrier density operator $\hat n (r)$ and the matrix element of carrier--phonon coupling for the phonon mode $\omega_{qj}$ with distinct polarization, respectively. 
(See Appendix for further discussion on various matrix element $M_{q,\nu}^{(j)}$ used in our computations.) 
Here $q$ and $k$ denote vectors $\vec q$ and $\vec k$, respectively, and $c_{k\nu}$ ($c_{k\nu}^\dagger$) is the ordinary annihilation (creation) operator of a carrier in the band $\nu$ and $a_{q}^{(j)}$ ($a_{q}^{(j)\dagger}$), the annihilation (creation) operator for a phonon with frequency $\omega_{qj}$ \cite{Fetter}.
Carrier--phonon interaction induces transition $|i\rangle \rightarrow |f\rangle$ of energy $E_i=E_i^{\rm c}+E_i^{\rm ph}$ and $E_f=E_f^{\rm c}+E_f^{\rm ph}$.
We consider the case that carriers (electrons and holes) and the lattice are weakly coupled so that carrier--phonon interaction $H_{\rm c-ph}^{(j)}$ would be treated as a perturbation. 
       
\subsection{Dielectric response functions}

The carrier-carrier interaction introduces dielectric screening and would weaken the carrier-phonon coupling.
As an illustration of linear response of a mcp, we consider an electron--hole plasma, a two-component plasma (2cp), subject to a weak external potential field $\phi_{\rm ext}(q,\omega)$ caused by some external test charge distribution $\rho_{\rm ext}$.
Electrons and holes respond differently to an external disturbance resulting in nonidentical  polarization functions $\Pi_{\nu\nu}^{\rm rpa} (q,\omega)$ ($\nu= \rm e ~ or ~ hh$).
The potential field $\phi_{\rm ext}(q,\omega)$ gives rise to corresponding potential energy $V_{\rm ext}^i$ for a carrier of type $i$. 
Here $i$ denotes different species of the plasma and we limit our consideration to the simplest case of a nondegenerate valence band in order to simplify our discussion, i.e., $i=1$ (2) for electrons (holes) of the plasma. 
We note that $V_{\rm ext}^i(q,\omega)=e_i \phi_{\rm ext}(q,\omega)$ for the carriers of electric charge $e_i=\mp e$, where $-e$ is the elementary charge of an electron. 

The external potential will cause changes in densities of each plasma species, and the charge density fluctuation $\delta \rho_i$ in the $i^{\rm th}$ component is written as
\be
\delta \rho_i(q,\omega) = \sum_{ij}\Pi_{ij}(q,\omega)V_{\rm ext}^j(q,\omega), \label{lrt}
\ee
where $\{V_{\rm ext}^1, V_{\rm ext}^2\}$ are the carrier density probe and $\Pi_{ij}(q,\omega)$ is the reducible polarization propagator -- the Fourier transform of the density-density response function of a mcp.
That is, $\mathcal{R}e~ \Pi_{ij}(q,\omega)$ is a measure of the response of an electron liquid to a bare external disturbance.
The imaginary part of the retarded polarization function, $\mathcal{I}m~ \Pi_{ij} (q,\omega)$, is directly linked to the real part of the conductivity -- a measure of dissipative processes, in which quanta of wave number $q$ and frequency $\omega$ are absorbed by the carriers in the plasma.
Now, the density fluctuation $\delta \rho_i(q,\omega)$ introduces polarization field to the carriers in the plasma, and the effective (self-consistent) potential energy of a carrier in the $i^{\rm th}$ plasma component is given by
\be
V_{\rm et}^i(q,\omega)=V_{\rm ext}^i(q,\omega)+\sum_j \psi_{ij}(q,\omega)\delta \rho_j(q,\omega). \label{Veff}
\ee 
Here the second term on the right hand side denotes the additional potential due to the polarization of the system, and $\psi_{ij}$ ($\neq \psi_{ji}$ in general) is the effective interaction between carriers of the $i^{\rm th}$ and $j^{\rm th}$ components including exchange and correlation effects.
In general, $\psi_{ij}$ is nonlocal and need be determined self-consistently because exact expression for $\psi_{ij}$ is not available.
Within the local density functional scheme of Kohn and Sham \cite{Kohn-Sham1965,KukkonenOverhauser1979}, $\psi_{ij}$ can be represented as $\psi_{ij}=v_{ij}[1-G_{ij}(q)]$ with $v_{ij}=\frac{4\pi e_i e_j}{\kappa q^2}$ and generalized local-field corrections $G_{ij}(q)$ as of Hubbard \cite{Hubbard,Vashishta1974}. 
Here $\kappa$ is the background dielectric constant of the material.   
Our discussion on the exchange--correlation effect of carriers parallels that of Kukkonen and Overhauser \cite{KukkonenOverhauser1979} and Hedin and Lundqvist \cite{HedinLundqvist1971} but with extension to the case of mcp \cite{Yi-Quinn1996}.
(For our convenience, we will minimize reminding the wave-vector and frequency dependences of the quantities explicitly in each expression now on. For example, $\Pi_{ij}(q,\omega)$ will be written as $\Pi_{ij}$.)   

The density fluctuation $\delta \rho_i$ can also be determined by treating the plasma as a noninteracting system responding to the self-consistent effective potential $V_{\rm et}^i$ such as
\be
\delta \rho_i(q,\omega) = \tilde{\Pi}_{i}(q,\omega)V_{\rm et}^i(q,\omega). \label{rpa}
\ee 
Here $\tilde{\Pi}_{i}$ denotes the temperature-dependent proper polarization function of the multi-component many carrier system and describes the response of a plasma component of type $i$ to the effective potential $V_{\rm et}^i$, the sum of the external and polarization potentials.
Hence $\tilde{\Pi}_{i}$ is a measure of the density fluctuation induced by a screened external charge $\tilde{\rho}_{\rm ext}^i (=\rho_{\rm ext}^i /\tilde{\varepsilon})$. 

Combining Eqs.(\ref{Veff}) and (\ref{rpa}) gives rise to coupled equations of $\delta \rho_i$ and $\delta \rho_j$.
One can easily solve the equations for $\delta \rho_i$ and $\delta \rho_j$ in terms of $V_{\rm ext}^i$ and $V_{\rm ext}^j$ to write
\be
\left( \begin{array}{c}
\delta \rho_1(q,\omega) \\ \delta \rho_2(q,\omega)
\end{array}
\right)
= \left( \begin{array}{cc}
\Pi_{11} & \Pi_{12} \\
\Pi_{21} & \Pi_{22} 
\end{array} \right)
\left(\begin{array}{c}
V_{\rm ext}^1(q,\omega) \\ V_{\rm ext}^2(q,\omega)
\end{array}  \right),     \label{coupled delta rho}
\ee
where individual components of the polarization propagators $\Pi_{ij}(q,\omega)$ are now given by
\be
\Pi_{ii}={\tilde{\Pi}_{i}(1- \tilde{\Pi}_{j}\psi_{jj})}/{\Delta}\mbox{  } \label{Piii}
\ee
for the intra-species  (diagonal) parts [$(i,j)=(1 \mbox{ or } 2)$] and
\be
\Pi_{ij}={\tilde{\Pi}_{i}\tilde{\Pi}_{j}\psi_{ij}}/{\Delta}\mbox{  } \label{Piij}
\ee
for the inter-species (off-diagonal) parts ($i\neq j$).
Here $\Delta$ is defined by $\Delta=1-\tilde{\Pi}_{1}\psi_{11} -\tilde{\Pi}_{2}\psi_{22} +\tilde{\Pi}_{1}\tilde{\Pi}_{2}(\psi_{11}\psi_{22}-\psi_{12}\psi_{21})$ and plays the role of the effective dielectric function in a mcp. 
(See Eq. (\ref{eff epsiloninv}) below.)

In the mean-field rpa, the proper polarization function $\tilde{\Pi}_{i}$ is replaced with the noninteracting expression commonly known as Lindhard polarizability ${\Pi}_{ii}^0$($\equiv\Pi_{i}^{0}$ for simplicity of notation).
It means that the effects of many-body correlations are neglected to let the effective pair interaction $\psi_{ij}$ become the bare Coulomb interaction $v_{ij}(=v_{ji})=\mp v$ with $v=4\pi e^2/(\kappa q^2)$.
The generalized noninteracting polarization function $\Pi_{ij}^{0}$ is given by 
\cite{Mahan,KimParkYi2011}
\be
\Pi_{ij}^0(q,\omega) = 2\sum_{k}\frac{f_{k+q,i}^{(0)}-f_{k,j}^{(0)}}{\varepsilon_{k+q,i}-\varepsilon_{k,j}-\hbar\omega-i\eta}, 
\label{Pi0}
\ee
where $f_{k,j}^{(0)}$ is the temperature dependent carrier distribution function.
The Lindhard-type expression of Eq. (\ref{Pi0}) is analogous to the case of spin-resolved expression in a multi-component spin system \cite{Yi-Quinn1996}.
The intra and inter-species components of the polarization propagator ${\Pi}_{ij}$ are written, in the rpa, as 
\be
\Pi_{ii}^{\rm rpa}=\Pi_{i}^0(1- v\Pi_{j}^0) /\left[1- v(\Pi_{i}^0+\Pi_{j}^0)\right] \label{Piii rpa}
\ee
and 
\be
\Pi_{ij}^{\rm rpa}=-v\Pi_{i}^0 \Pi_{j}^0 /\left[1- v(\Pi_{i}^0+\Pi_{j}^0)\right]. \label{Piij rpa}
\ee
We note that $\Delta^{\rm rpa}=1- v(\Pi_{1}^0+\Pi_{2}^0)$ is the effective macroscopic dielectric function $\epsilon_{\rm eff}^{\rm rpa}$ in a mcp. (See further discussion on $\epsilon_{\rm eff}$ below.)
Further reducing to the case of single-component system such as in doped semiconductors, we have $\Pi_{ii}^{\rm rpa}=\Pi_{i}^0/(1- v\Pi_i^0)$, the well known expression of mean-field polarizability \cite{Mahan}. 
Since $\Pi_{ii}^0(q,\omega,T)$ is a complex function, the real and imaginary parts $ \mathcal{R}e ~\Pi_{ii}(q,\omega,T)$ and $\mathcal{I}m ~\Pi_{ii}(q,\omega,T)$ 
can be expressed in terms of $\mathcal{R}e ~\Pi^0_{ii}$ and $\mathcal{I}m ~\Pi^0_{ii}$, which are given, respectively, by \cite{Maldague,Giuliani}
\begin{align}
\mathcal{R}e ~\Pi_{ii}^0 (q,\omega;T) =& -g_{i} \int_0^\infty dx \frac{F(x,T)}{q/k_{Fi}}  \nonumber\\
\times 
\frac{1}{2}&\left[\ln \left|\frac{x-\eta_{i-}}{x+\eta_{i-}}\right|-\ln \left|\frac{x-\eta_{i+}}{x+\eta_{i+}}\right| \right]
\label{Re Pi0}
\end{align}
and
\begin{align}
\mathcal{I}m ~\Pi_{ii}^0 &(q,\omega;T) \nonumber \\
=-\pi g_{i} &\left[\frac{\omega}{v_{Fi}  q}
+\frac{k_{\rm B} T}{\hbar v_{Fi}  q}
\ln \frac{1+e^{\beta[\eta_{i-}^2\varepsilon_{Fi}  - \mu_{i} (T)]}}{1+e^{\beta[\eta_{i+}^2
\varepsilon_{Fi}  - \mu_{i}(T)]}}\right].
\label{Im Pi0}
\end{align} 
Here $\beta=1/k_{\rm B}T$, $F(x,T)=x[e^{\beta\left(x^2\varepsilon_{Fi} - \mu_{i}(T)\right)}+1]^{-1}$, 
$g_{i}=\frac{m_{i}k_{Fi}}{\pi^2\hbar^2}$, $k_{Fi}=(3\pi^2 n_i)^{1/3}$,
and $\eta_{i\pm}=\frac{\omega}{qv_{Fi}} \pm\frac{q}{2k_{Fi}}$.
In Eqs.(\ref{Re Pi0}) and (\ref{Im Pi0}), $T$ is the effective temperature of the corresponding carriers in quasi-equilibrium and conventional notations are used such as $m_{i}$, $k_{Fi}$, $v_{Fi}$, $\varepsilon_{Fi}$, and $\mu_{i}$ denoting effective mass, Fermi wave number, Fermi velocity, Fermi energy, and chemical potential, respectively, of a carrier (either electrons or holes) indicated by subscript $i$. 
\begin{figure*}[ht]  
\includegraphics*[width=.9\textwidth]{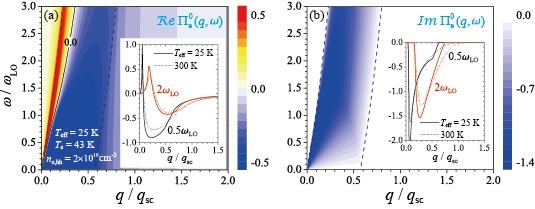}
\caption{
(Color Online) Real and imaginary parts of noninteracting polarization functions $\Pi_c^0 (q,\omega)$ of conduction electron plasma for the carrier density of $2 \times 10^{19} \rm cm^{-3}$ at effective temperature $T_{\rm eff}$=25 K.
Insets illustrate the wave-number dependence of $\Pi_c^0 (q,\omega)$ for representative frequencies at 25 K and 300 K.
Pair of dashed lines denotes the region of allowed single-particle excitations in a wurtzite GaN.}
\label{P0_ri_wq_25K}
\end{figure*}
In Fig.\ref{P0_ri_wq_25K} real and imaginary parts of $\Pi_{ii}^0 (q,\omega)$ ($i=e$) of a conduction electron plasma is illustrated in the $\omega$--$q$ plane for the carrier density of $2 \times 10^{19} \rm cm^{-3}$ at effective temperature $T_{\rm eff}$=25 K.
Each inset illustrates the wave number dependence of $\mathcal{R}e ~\Pi_{ii}^0 (q,\omega)$ and $\mathcal{I}m ~\Pi_{ii}^0 (q,\omega)$, respectively, for representative frequencies at 25 K and 300 K. 
Carriers within the Fermi sea can be excited to states outside the Fermi sea, and single-particle excitations of free electrons or free holes are the only processes for the energy and momentum dissipation, since the effects of carrier screening is completely neglected in $\Pi_{ii}^0 (q,\omega)$ \cite{Yi-Kim-CAP2015}. 
Real part of the noninteracting polarization function $\mathcal{R}e ~\Pi_{ii}^0 (q,\omega)$ shows a broad valley structure within the single-particle excitation continuum at small frequency $\omega \ll \omega_{\rm LO}$ for $q < q_{\rm sc}$ and a peak structure of moderate width along the upper boundary of the continuum.
The line of $\mathcal{R}e ~\Pi_{ii}^0 (q,\omega)=0$ lies in the continuum region of the single-particle excitations and, at small $q$, $\mathcal{R}e ~\Pi_{ii}^0 (q,\omega)$ changes sign from negative to positive as $\omega$ increases sweeping across the continuum region.
On the other hand, $\mathcal{I}m ~\Pi_{ii}^0 (q,\omega)$ is finite and negative inside the continuum region of the single-particle excitations showing a sharp dipped structure along the zero line of  $\mathcal{R}e ~\Pi_{ii}^0 (q,\omega)$.

Since we now have all the components of the polarization propagator $\Pi_{ij}$, the self-consistent interaction of Eq.(\ref{Veff}) can now be written, with $\delta \rho_i$'s given by Eq.(\ref{coupled delta rho}), as
\be
\left( \begin{array}{c}
V_{\rm et}^1 \\ V_{\rm et}^2
\end{array}
\right)
&=& \frac{1}{\Delta}\left( \begin{array}{cc}
1-\Pi_2^0\psi_{22} & \Pi_2^0\psi_{12} \\
\Pi_1^0\psi_{21} & 1-\Pi_1^0\psi_{11} 
\end{array} \right)
\left(\begin{array}{c}
V_{\rm ext}^1 \\ V_{\rm ext}^2    \nonumber
\end{array}  \right) \\
&\equiv&
\left( \begin{array}{cc}
\tilde\epsilon_{11}^{-1} & \tilde\epsilon_{12}^{-1} \\
\tilde\epsilon_{21}^{-1} & \tilde\epsilon_{22}^{-1} 
\end{array} \right)
\left(\begin{array}{c}
V_{\rm ext}^1 \\ V_{\rm ext}^2
\end{array}  \right).     \label{coupled Veff}
\ee
Here the $2 \times 2$ matrix multiplied to the column of external probe potentials $(V_{\rm ext}^1, V_{\rm ext}^2)$ on the right hand side is just the inverse of `plasma--test charge' dielectric tensor ${\underline{\tilde\epsilon}}$ in the $2\times 2$ space of plasma species and each component is given by
\be
\tilde\epsilon_{ij}^{-1} =\delta_{ij} + \sum_{\ell=1,2}\psi_{i\ell}\Pi_{\ell j}. \label{et epsilon}
\ee
(Similar description corresponding to the case of spin-polarized electrons was investigated earlier by one of us \cite{KimParkYi2011}.)
The intra-species `electron--test charge' dielectric function $\tilde\epsilon_{ii}=1-\Pi_i^0\psi_{ii}-\Pi_i^0\psi_{ij}\Pi_j^0\psi_{ji}/(1-\Pi_j^0\psi_{jj})$ reduces, in the rpa, to  $\tilde\epsilon_{ii}^{\rm rpa}=1-v\Pi_i^0/(1-v\Pi_j^0)$ and to $\tilde\epsilon^{\rm rpa}=1-v\Pi_i^0$ in the case of single species electron liquid \cite{HedinLundqvist1971}.
On the other hand, the inter-species component $\tilde\epsilon_{ij}=\Delta/(\psi_{ij}\Pi_j^0)$ reduces to $\tilde\epsilon_{ij}^{\rm rpa}=1-(1-v\Pi_i^0)/(v\Pi_j^0)$ in the rpa and is undefined in the case of single-species electron system.
We find that 
\be
\sum_{ij} (\tilde\epsilon_{ij}^{-1}-&\delta_{ij})=[(\psi_{11}+\psi_{21})\Pi_1^0 +(\psi_{22}+\psi_{12})\Pi_2^0  \nonumber \\
&-2(\psi_{11}\psi_{22}-\psi_{12}\psi_{21})\Pi_1^0\Pi_2^0]/\Delta  \label{sum rule1}
\ee   
and, hence, that $\sum_{ij} (\tilde\epsilon_{ij}^{-1}-\delta_{ij})|_{\rm rpa} =0$ since $\psi_{ii}=-\psi_{ij} (i\neq j)$ in the rpa.

In response to the potential field $\phi_{\rm ext}$ due to an external test charge $\rho_{\rm ext}^i$ ($\rho_{\rm ext}^j$) of electrical charge $e_i$ ($e_j$), another probing test charge would experience the `test charge--test charge' interaction $V_{\rm tt}^i$ ($V_{\rm tt}^j$) written as
\be
V_{\rm tt}^{i}(q,\omega)= v_{ii}(\rho_{\rm ext}^{i}+\delta\rho_{i} ) +v_{ij}\delta\rho_{j}, \label{vtt}
\ee 
and, similarly, $V_{\rm tt}^{j}(q,\omega)$ with the indices $i$ and $j$ interchanged in $V_{\rm tt}^{i}$. 
In general, the dielectric function $\epsilon(q,\omega)$ of a material is defined, in terms of `test charge--test charge' interaction, by
$V_{\rm tt}(q,\omega)= V_{\rm ext} (q,\omega)/\epsilon(q,\omega)$ \cite{KukkonenOverhauser1979}, and can be extended to a mcp as follows.
On substituting the density fluctuations $\delta\rho_i$'s of Eq.(\ref{coupled delta rho}) into Eq.(\ref{vtt}), $V_{\rm tt}^i$ and $V_{\rm tt}^j$ are written, in terms of $V_{\rm ext}^i$ and $V_{\rm ext}^j$, as
\be
\left( \begin{array}{c}
V_{\rm tt}^1 \\ V_{\rm tt}^2
\end{array}
\right)
&=& \left( \begin{array}{cc}
\epsilon_{11}^{-1} & \epsilon_{12}^{-1} \\
\epsilon_{21}^{-1} & \epsilon_{22}^{-1} 
\end{array} \right)
\left(\begin{array}{c}
V_{\rm ext}^1 \\ V_{\rm ext}^2   
\end{array}  \right).  \label{coupled Vtt}
\ee
Here $\epsilon_{ij}^{-1}$'s are the components of the inverse dielectric tensor ${\underline{\epsilon}}^{-1}$ of a mcp and are given in terms of the polarization propagators.
The intra- and inter-species components are expressed, respectively, as $\epsilon_{ii}^{-1}=1+v(\Pi_{ii}-\Pi_{ji})$ and 
$
\epsilon_{ij}^{-1}=v(\Pi_{ij}-\Pi_{jj})
$
and can be combined to be written, in general, as
\be
 \epsilon_{ij}^{-1} =\delta_{ij} + (-1)^{i+j}\sum_{\ell=1,2}v_{j\ell}\Pi_{\ell j}.\label{tt epsilon}
\ee
On substituting  $\Pi_{ij}$'s of Eqs.(\ref{Piii}) and (\ref{Piij}) into Eq.(\ref{tt epsilon}), one can obtain, in a mcp, an identity of
\be
\sum_{ij} (\epsilon_{ij}^{-1}-\delta_{ij}) =0.    \label{sum rule2}  
\ee  
In the rpa, $V_{\rm et}^{j}=V_{\rm tt}^{j}$ and, hence, $\tilde\epsilon_{ij}$ and $\epsilon_{ij}$ are identical.
If we neglect inter-species correlations (i.e., $G_{ij}=G\delta_{ij}$ as of Hubbard's local-field correction \cite{Hubbard}), we have 
\be
\epsilon_{ii}=1-\frac{\Lambda_i v \Pi_{i}^0}{1-\Lambda_j v\Pi_j^0}
\ee
and
\be
\epsilon_{ij}=1-\frac{1-\Lambda_i v \Pi_{i}^0}{\Lambda_j v\Pi_j^0}.
\ee
Here the vertex function $\Lambda_i$ is defined by $\Lambda_i =[1-(\psi_{ii}-v)\Pi_i^0]^{-1}$ generally known as $1/(1+vG_{ii}\Pi_{i}^0)$ \cite{Mahan}.
For the case of a single species system, one resumes $\epsilon_{ii}^{-1}=1+v\Pi_{i}^0/(1-\psi_{ii}\Pi_{i}^0)$ giving rise to $\epsilon_{ii}=1-\Lambda_i v\Pi_{i}^0$.
Since $\Lambda_i\rightarrow 1$ in the rpa, we have $\epsilon_{ij}^{\rm rpa}=1-{(1-v \Pi_{i}^0)}/{(v\Pi_j^0)}$ and $\epsilon_{ii}^{\rm rpa} =1-v\Pi_{i}^0/(1-v\Pi_{j}^0)$ further reducing to $\epsilon_{ii}^{\rm rpa}=1-v\Pi_{i}^0$ in a single species electron gas.

\begin{figure*}[t] 
\includegraphics*[width=.8\textwidth]{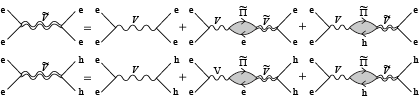}
\caption{Dyson equations for the effective Coulomb interaction $\tilde{V}_{ij}$ ($i=e {\rm ~or~} h$) between carriers in a multi-component plasma}
\label{Vdyson}
\end{figure*}
Let us introduce the macroscopic effective dielectric constant $\epsilon_{\rm eff}$ of a mcp by equating $[\epsilon_{\rm eff}^{-1}(q,\omega) -1]\rho_{\rm ext}$ phenomenologically in the presence of the external test charge $\rho_{\rm ext}$ to the net charge density induced in the plasma as follows \cite{Pines}
\be
[\epsilon_{\rm eff}^{-1}(q,\omega) -1]\rho_{\rm ext}\equiv\delta\rho_1(q,\omega)+\delta\rho_2(q,\omega). \label{eff epsilon-1} 
\ee
Substitution of $\delta\rho_i$ given by Eq.(\ref{lrt}) into Eq.(\ref{eff epsilon-1}) along with Eqs.(\ref{Piii}) and (\ref{Piij}) leads us to
\be
\epsilon_{\rm eff}^{-1}(q,\omega)-1=v[(\Pi_1^0+\Pi_2^0-\Pi_1^0\Pi_2^0\sum_{ij=1,2}\psi_{ij}]/\Delta ~~~    \label{eff epsiloninv} \\
=(\epsilon_{11}^{-1}-1) +(\epsilon_{22}^{-1}-1) =-(\epsilon_{12}^{-1}+\epsilon_{21}^{-1}), \nonumber
\ee
where the identity of Eq.(\ref{sum rule2}) has been observed in writing the last equality.
In the rpa, $\epsilon_{\rm eff}^{\rm rpa}=1-v(\Pi_1^0+\Pi_2^0)$, which is the same as that suggested by Vashishta et al. and others \cite{Vashishta1974,Collet1989}. 
However, we note that Vashishta et al. introduced their effective dielectric constant $\epsilon_{\rm eff}$ differently. 
(They defined their $\epsilon_{\rm eff}^{-1}(q,\omega) -1$ as $\sum_{ij} (\epsilon_{ij}^{-1}-\delta_{ij})$,  which vanishes according to the identity given by Eq.(\ref{sum rule2}).)    
Our description of $\epsilon_{\rm eff}$ is consistent with that of the Dyson equation approach.
\begin{figure*}[t]
\includegraphics[width=.8\textwidth]{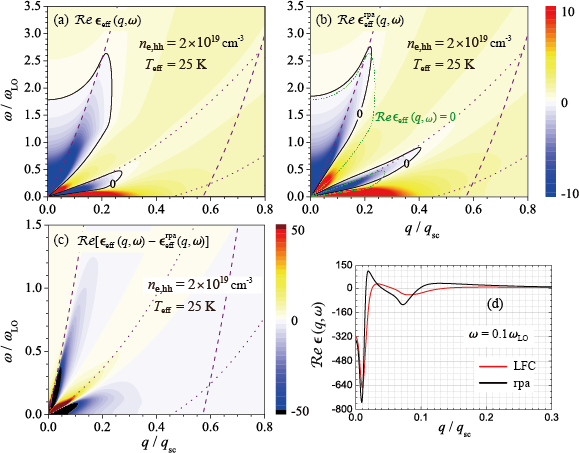}
\caption{(Color Online) Effective dielectric functions $\mathcal{R}e ~\epsilon_{\rm eff}(q,\omega)$ of two-component plasma formed in a wurtzite GaN with conduction electron density of $2 \times 10^{19} \rm cm^{-3}$ at effective carrier temperature 25 K: (a) $\mathcal{R}e ~\epsilon_{\rm eff}(q,\omega)$. (b) $\mathcal{R}e ~\epsilon_{\rm eff}^{\rm rpa}(q,\omega)$. 
The contour of $\mathcal{R}e ~\epsilon_{\rm eff}(q,\omega)=0$ is denoted in green dotted line for comparison.
(c) $\mathcal{R}e ~[\epsilon_{\rm eff}(q,\omega)-\epsilon_{\rm eff}^{\rm rpa}(q,\omega)]$, and (d) $\mathcal{R}e ~\epsilon_{\rm eff}(q,\omega)$ and $\mathcal{R}e ~\epsilon_{\rm eff}^{\rm rpa}(q,\omega)$ for $\omega = 0.1 \omega_{\rm LO}$. 
Pairs of dashed lines denote the corresponding boundaries of allowed single-particle excitation continuum.}
\label{epsilon}
\vspace{1.0cm}
\includegraphics*[width=.8\textwidth]{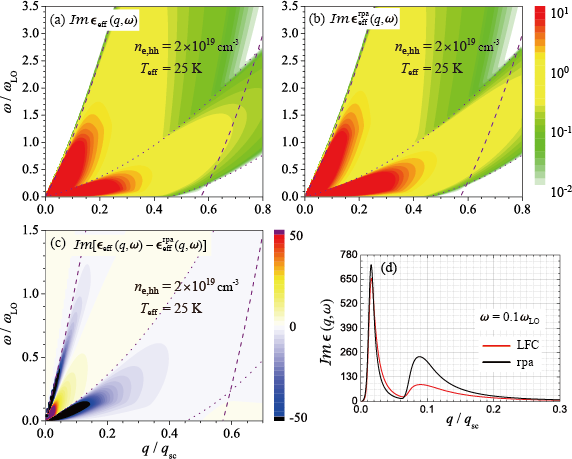}
\caption{(Color Online) Effective dielectric functions $\mathcal{I}m ~\epsilon_{\rm eff}(q,\omega)$ of two-component plasma formed in a wurtzite GaN with conduction electron density of $2 \times 10^{19} \rm cm^{-3}$ at effective carrier temperature 25 K: (a) $\mathcal{I}m ~\epsilon_{\rm eff}(q,\omega)$. (b) $\mathcal{I}m ~\epsilon_{\rm eff}^{\rm rpa}(q,\omega)$. 
(c) $\mathcal{I}m~[\epsilon_{\rm eff}(q,\omega)-\epsilon_{\rm eff}^{\rm rpa}(q,\omega)]$, and (d) $\mathcal{I}m~\epsilon_{\rm eff}(q,\omega)$ and $\mathcal{I}m ~\epsilon_{\rm eff}^{\rm rpa}(q,\omega)$ for $\omega = 0.1 \omega_{\rm LO}$. 
Pairs of dashed lines denote the corresponding boundaries of allowed single-particle excitation continuum.}
\label{epsilon_im}
\end{figure*} 
In a multi-component many carrier system (consisting of electrons and various holes in the present case), the effective (dressed) interactions $\tilde{V}_{ij}$ between carriers of types $i$ and $j$ are the solutions of Dyson equations.
For example, in terms of full retarded (proper) polarization function $\tilde{\Pi}_{\ell}$ and bare Coulomb interaction $v_{ij}$, it is written as
\be
\tilde{V}_{ij}=v_{ij}+ \sum_{\ell=e,h} v_{i\ell}\tilde{\Pi}_{\ell}\tilde{V}_{\ell j}.
\label{V dyson}
\ee
Here the bare Coulomb interaction $v_{ij}$ is $v_q (\equiv \frac{4\pi e^2}{q^2})$ for $i=j$ or $-v_q$ for $i\neq j$.
The internal Coulomb interactions of the carriers in the plasma are renormalized in exactly the same way as the external potential fields.
Dyson equations for dressed Coulomb interactions $\tilde{V}_{ee}$ between conduction electrons (denoted by $e$'s) and $\tilde{V}_{eh}$ between an electron and a hole (denoted by $h$) are illustrated in Fig. \ref{Vdyson}. 
One can solve the coupled equations of Eq. (\ref{V dyson}) explicitly for $\tilde{V}_{ee}$, $\tilde{V}_{hh}$, and $\tilde{V}_{eh}$ to write $\tilde{V}_{ij}=v_{ij}/\epsilon_{\rm eff}$ confirming $\epsilon_{\rm eff}^{\rm rpa} (q,\omega) = 1-v_q \sum_{i=e,h} {\Pi}_{i}^0(q,\omega)$.

In Fig. \ref{epsilon}(a) and (b), dispersive behaviors of the real part of effective dielectric function $\epsilon_{\rm eff}(q,\omega)$ of Eq.(\ref{eff epsiloninv}) and $\epsilon_{\rm eff}^{\rm rpa}(q,\omega)$ are shown, respectively, for a 2cp formed of conduction electrons and heavy holes each with concentration $2 \times 10^{19} \rm cm^{-3}$ at effective carrier temperature 25 K. 
The zero value contours of $\mathcal{R}e ~\epsilon_{\rm eff}(q,\omega)=0$ are indicated with dark solid lines, each denoting the dispersion curves of high-frequency `optic' and low-frequency `acoustic' plasmon modes.
Pairs of dashed and dotted lines indicate the boundaries of allowed single-particle excitation continua for electrons and heavy holes.
In the region of long wavelength and high frequency, $\mathcal{I}m ~\Pi(q,\omega)$ vanishes (See Fig. \ref{PI_ri_wq_25K} below.) and, hence, $\mathcal{I}m ~\epsilon_{\rm eff}(q,\omega)=0$ allowing well-defined dissipationless self-sustaining collective oscillations.
The difference of $\mathcal{R}e ~[\epsilon_{\rm eff}(q,\omega)-\epsilon_{\rm eff}^{\rm rpa}(q,\omega)]$ is illustrated in panel (c), and the wave-number dependences of $\mathcal{R}e~\epsilon_{\rm eff}(q,\omega)$ and 
$\mathcal{R}e~\epsilon_{\rm eff}^{\rm rpa}(q,\omega)$ are compared in panel (d) for $\omega = 0.1 \omega_{\rm LO}$. 
In panel (b), the contour of $\mathcal{R}e ~\epsilon_{\rm eff}(q,\omega)=0$ is indicated by green dotted line for comparison.     
In Fig. \ref{epsilon}(a), we find that a pair of plasmon branches are observed in a 2cp, and that optic and acoustic branches are well separated within the rpa, each damped through single-particle excitations of electrons and holes, respectively. 
However, the local-field corrections of carriers modify the plasmon branches slightly reducing plasmon frequencies of both the optical and acoustical branches for given values of wave number $q$.  
We note that, in the 2cp of $n_e=2 \times 10^{19} \rm cm^{-3}$, the higher frequency electron plasmon mode of bare frequency $\omega_{\rm p,e}\simeq 1.8 \omega_{\rm LO}$ is almost intact and well defined but that the lower frequency plasmon mode of heavier species (heavy holes) at $\omega_{\rm p,hh} \simeq 0.69 \omega_{\rm LO}$ is screened by the lighter conduction electrons giving rise to an acoustic branch, the latter mode being subject to Landau damping by the lighter species, since the branch is located well inside the electron excitation continuum \cite{YI&Kim_PhyscaB}.
Dispersive behaviors of $\mathcal{I}m ~\epsilon_{\rm eff}(q,\omega)$ and $\mathcal{I}m ~\epsilon_{\rm eff}^{\rm rpa}(q,\omega)$ are shown in Fig. \ref{epsilon_im} for a 2cp of conduction electrons and heavy holes each with concentration $2 \times 10^{19} \rm cm^{-3}$ at effective carrier temperature 25 K.
While $\mathcal{I}m ~\epsilon_{\rm eff}(q,\omega)$ has a single peaked structure in a $\omega-q$ plane for a scp \cite{Yi-Kim-CAP2015},
it reveals double peaked structure each appearing inside the single-particle excitation continua of electrons and holes in a 2cp, respectively, right below the upper boundaries of each continuum in the region of low frequency and long wavelength. 
The effects of local-field corrections of the carriers are appreciable only in the region of low frequency and long wavelength as illustrated in panels (c) and (d) of Fig. \ref{epsilon_im}.  

\subsection{Phonon spectral functions}
The dielectric screening in many carrier system gives rise to renormalized electron--phonon coupling and thus to dressed phonon propagator modifying the phonon dispersion relations along with phonon spectral function.
In compound semiconductors, charge carriers couple to phonons via various channels such as couplings to LO and TO phonons, and also through acoustic deformation potential and piezoacoustic (AP) couplings. 
The interaction of specific phonon--carrier coupling (designated by a parameter $j$) represented by $H_{\rm c-ph}^{(j)}$ is given in Eq.(\ref{Heph}).  

The phonon spectral function $\mathcal{A}(q,\omega)$ of a material describes the probability distribution of having phonons with wave number $q$ and frequency $\omega$.
The bare phonon modes in a solid would be modified due to carrier screening and phonon--plasmon coupling, and the poles of the dressed phonon propagator determine the renormalized phonon dispersion relations, and
the effective phonon spectral function of a solid is, in general, the sum of contributions from each individual electron--phonon coupling channel such that  
\be
\mathcal{A}(q,\omega)=\sum_{j\nu} \mathcal{A}_{j\nu}(q,\omega).
\ee
Here $\mathcal{A}_{j\nu}(q,\omega)$ describes the probability distribution of having phonons (of $j^{\rm th}$ type) dressed by plasma species $\nu$ and is given, in terms of retarded phonon propagator $D_{j\nu} (q,\omega)$ for the individual phonon mode, as \cite{Mahan} 
\be
\mathcal{A}_{j\nu}(q,\omega)= -\frac{1}{\pi}\mathcal{I}m ~D_{j\nu} (q,\omega).
\ee

The dielectric screening in many carrier system gives rise to renormalized electron--phonon coupling and thus to dressed phonon propagator \cite{Hwang2007}. 
This collisional broadening modifies the phonon dispersion relations along with phonon spectral function.
The dressed phonon propagator $D_{j\nu}$ with collisional broadening is written, in general, as \cite{Bruus,Mahan72}
\begin{align}
D_{j\nu}(q,\omega)= & \frac{2\omega_{q j}}{\omega^2-  \omega_{q j}^2 - 2 \omega_{q j}\mid M_{q}^{(j)}\mid^2 \Pi_{\nu\nu}(q,\omega)/\hbar},
\label{dressed phonon propagator}
\end{align}
where $\omega_{q j}$ is the bare (undoped crystal) phonon frequency of mode $j$. 
Ignoring the phonon renormalization, $D_{j\nu}$ reduces, with an infinitesimal positive $\eta$, to \cite{Mahan} 
\begin{align}
D_j^{(0)}(q,\omega)= & \frac{2\omega_{q j}}{\omega^2-  \omega_{q j}^2 +i\eta},
\label{bare phonon propagator}
\end{align}
giving rise to the well-known bare phonon spectral function 
$\mathcal{A}_j^{(0)}(q,\omega)= [\delta(\omega-\omega_{q j})- \delta(\omega+\omega_{q j})]$.
In Eq.(\ref{dressed phonon propagator}), $\Pi_{\nu\nu}(q,\omega)$ is the full retarded polarization propagator of each plasma component (distinguished by $\nu$) given by Eq. (\ref{Piii}), and
$\mid M_{q,\nu}^{(j)}\mid^2 \Pi_{\nu\nu}(q,\omega)/\hbar$ in the denominator represents the (complex) phonon self-energy correction $\mathcal{P}_{j\nu}(q,\omega)(\equiv \Delta_{j\nu}-i\Gamma_{j\nu}/2)$ via polarization function of plasma species $\nu$, which is known to introduce low energy quantum interference branch in the phonon spectral function \cite{Jain1988}. 
The real and imaginary parts of the phonon self-energy, $\Delta_{j\nu}$ and $\Gamma_{j\nu}$, are given, respectively, 
by $\Delta_{j\nu}(q,\omega)= \mid M_{q,\nu}^{(j)}\mid^2\mathcal{R}e ~\Pi_{\nu\nu}(q,\omega)/\hbar$ and $\Gamma_{j\nu}(q,\omega)=-2\mid M_{q}^{(j)}\mid^2\mathcal{I}m~\Pi_{\nu\nu}(q,\omega)/\hbar$, 
the former describing the frequency renormalization correction due to the electronic screening of the long-ranged Coulomb fields associated with the phonons and the latter being a measure of phonon lifetime $\tau_{q j}$ or the width of the spectral function due to collisional broadening. 
Phonon self-energy and, hence, $\Delta_{j\nu}$ and $\Gamma_{j\nu}$ are also functions of $\omega$ and $q$, and $\mathcal{I}m~\Pi_{\nu\nu}(q,\omega) \le 0$, in general.

Now, the phonon spectral function $\mathcal{A}_{j\nu}(q,\omega)$ is written as
\begin{align}
\mathcal{A}_{j\nu}& (q,\omega)  \nonumber \\
= &\frac{\omega_{q j}^2 \Gamma_{j\nu}(q,\omega)}{[\omega^2-  \omega_{q j}^2 -2 \omega_{q j}\Delta_{j\nu}(q,\omega)]^2 + [\omega_{q j}\Gamma_{j\nu} (q,\omega)]^2 }.
\label{Ajqw}
\end{align}
The denominator of $\mathcal{A}_{j\nu}(q,\omega)$ can be rewritten, in terms of phenomenological renormalized phonon frequency $\tilde{\omega}_{q j}$ and phonon lifetime $\tau_{q j}$, as 
$\omega^2- (\tilde{\omega}_{q j} -\frac{i}{2\tau_{q j}})^2 $,
where $\tilde{\omega}_{q j}$ satisfies a quadratic equation given by
\be
\tilde{\omega}_{q j}^4 - \omega_{q j}^2 (1+{ 2 \Delta_{j\nu}}/{\omega_{q j}})\tilde{\omega}_{q j}^2 -\omega_{q j}^2\Gamma_{j\nu}^2 =0
\label{ren omega_ph}
\ee
with
$
\tau_{q j}^{-1}=\frac{\omega_{q j}}{\tilde{\omega}_{q j}} \Gamma_{j\nu}$. 
For each phonon mode $j$, the electron--phonon interaction introduces the phonon self-energy to change the bare phonon frequencies $\omega_{q j}$ to the new frequencies $\tilde{\omega}_{q j}$ with finite lifetime $\tau_{q j}$.
Equation (\ref{ren omega_ph}) gives rise to, along with the renormalized primary mode close to  
${\tilde\omega}_{j\nu}^+(q,\omega)\simeq \omega_{q j} (1+\frac{ 2 \Delta_{j\nu}}{\omega_{q j}}+\frac{\Gamma_{j\nu}^2/\omega_{q j}^2}{1+ 2 \Delta_{j\nu}/\omega_{q j}})^{1/2}$ 
for $ 2 \Delta_{j\nu} \ge -{\omega_{q j}}$, a secondary (low energy) mode ${\tilde\omega}_{j\nu}^-(q,\omega)\simeq \frac{\Gamma_{j\nu}}{{\sqrt{|1+\frac{ 2 \Delta_{j\nu}}{\omega_{q j}}|}}}$ for $2\Delta_{j\nu} < -\omega_{q j}$ with negative self-energy correction $\Delta_{j\nu} (<0)$.
We find that the latter mode ${\tilde\omega}_{j\nu}^-$ would be well-defined only with finite values of $\Gamma_{j\nu}(q,\omega)$ in the region of $\mathcal{R}e ~\Pi_{\nu\nu}(q,\omega)<0$ in the $\omega-q$ space. 
$\Gamma_{j\nu}(q,\omega)$ is finite only in the $\omega-q$ plane of finite $\mathcal{I}m ~\Pi_{\nu\nu}(q,\omega)$, which occurs in the presence of single-particle excitations as is shown in Fig. \ref{PI_ri_wq_25K} below.   
Deep valley with negative $\mathcal{R}e ~\Pi_{\nu\nu}(q,\omega)$ occurs in the (very) low energy region of high damping inside the single-particle excitation continuum on the $\omega-q$ plane.(See Fig. \ref{PI_ri_wq_25K}(a) and (b).)


\section{Results and Discussion}
In order to illustrate numerical results for the spectral behaviors of dielectric response and phonon spectral functions in  an ideal mcp, use has been made of effective masses $m_{\rm e}= 0.22 m_0$, $m_{\rm hh}= 1.3 m_0$, and $m_{\rm lh}= 0.30 m_0$ of a (simplified) wurtzite GaN with parabolic bands.

The bare plasma frequencies for carrier concentration of $2\times 10^{19} \rm cm^{-3}$ are $\omega_{\rm p,e}(\epsilon_\infty) \simeq 153 ~\rm meV ~(= 1.68\omega_{\rm LO})$, $\omega_{\rm p,hh}(\epsilon_\infty) \simeq 63 ~ \rm meV ~(=0.69 \omega_{\rm LO})$, and $\omega_{\rm p,lh}(\epsilon_\infty) \simeq 131~ \rm meV ~(= 1.44 \omega_{\rm LO})$ for scp plasmas of electrons, heavy holes, and light holes, respectively.
In the present work, frequencies and wave numbers are scaled by the longitudinal bare phonon frequency $\omega_{\rm LO}$ and the Thomas--Fermi screening wave number $q_{\rm sc}$, respectively.
We consider a simplified nondispersive model of Einstein for bare optical phonons with $\omega_{\rm LO}=92~ \rm meV$ and $\omega_{\rm TO}=66~ \rm meV (\simeq 0.72 \omega_{\rm LO})$ and a Debye-type model for bare acoustical phonons of $\omega_{q j}=s q$ in a undoped GaN. 
Here $s$ is the speed of sound wave in the material.
For a scp, $q_{\rm sc}$ is given, in terms of  particle number density $n$ and the chemical potential $\mu$, by 
$
q_{\rm sc}^2 = \frac{4\pi e^2}{\epsilon_{\infty}}\frac{\partial n}{\partial \mu}
$ \cite{ashcroft-mermin}, 
and we have $q_{\rm sc}^{(\rm e)}=9.2 \times 10^6 \rm cm^{-1}$ and $q_{\rm sc}^{(\rm hh)}=2.2 \times 10^7 \rm cm^{-1}$ at 25 K and $q_{\rm sc}^{(\rm e)}=8.6 \times 10^6 \rm cm^{-1}$ and $q_{\rm sc}^{(\rm hh)}=2.1 \times 10^7 \rm cm^{-1}$ at 300 K, respectively, for carrier concentration of $2 \times 10^{19} \rm cm^{-3}$.
For a mcp, $q_{\rm sc}$ is written as \cite{YI&Kim_PhyscaB}
$
q_{\rm sc}^2 = \frac{4\pi e^2}{\epsilon_{\infty}}\sum_{\nu}\frac{\partial n_\nu}{\partial \mu_\nu},
$
where $n_\nu$ and $\mu_\nu$ are, respectively, the particle number density and the quasi chemical potential of the band occupied by the plasma species $\nu$. 
For a 2cp consisting (of conduction electrons and heavy holes) with electron concentration of $2 \times 10^{19} \rm cm^{-3}$, we have $q_{\rm sc}=2.94 \times 10^7 \rm cm^{-1}$ at 25 K and $q_{\rm sc}=2.54 \times 10^7 \rm cm^{-1}$ at 300 K, respectively.

\begin{figure*}[th]  
\includegraphics*[width=.9\textwidth]{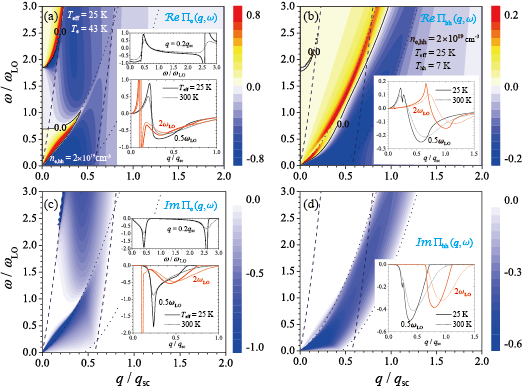}
\caption{(Color Online) Real and imaginary parts of dressed polarization functions, $\Pi_{\nu\nu} (q,\omega)$, for conduction electrons ($\nu= \rm e$) and heavy holes ($\nu= \rm hh$) of concentration $2 \times 10^{19} \rm cm^{-3}$ and effective carrier temperature 25 K. Insets illustrate the frequency or wave-number dependences of $\mathcal{R}e ~\Pi_{\nu\nu} (q,\omega)$ for representative values of wave number or frequencies.
Pairs of dashed and dotted lines denote the allowed regions of single-particle excitations for conduction electrons and heavy holes, respectively, in a wurtzite GaN.}
\label{PI_ri_wq_25K}
\end{figure*}
For a 2pc consisting of conduction electrons and heavy holes, each species of carrier concentration $2 \times 10^{19} \rm cm^{-3}$, the spectral behavior of the dressed polarization function $\Pi_{\nu\nu}^{\rm rpa} (q,\omega)$ ($\nu= \rm e ~ or ~ hh$) shown in Fig. \ref{PI_ri_wq_25K} reveals the multi-component character of the high-frequency optic and low-frequency acoustic plasmon branches.  
Boundaries of a pair of single-particle excitation continua for electrons and heavy holes are indicated with steeper dashed (for electrons) and slower dotted (for holes) lines, respectively, and branches of the optic and acoustic plasmon excitations are clearly distinguished in strong color intensities.
Each inset illustrates the frequency or wave-number dependence of $\Pi_{\nu\nu} (q,\omega)$ at 25 K and 300 K for representative wave number or frequencies, respectively.
In a mcp, in addition to the contribution from the conventional optic plasmon-LO phonon coupling of the scp such as of doped semiconductors \cite{DasSarma1990}, the contribution from the low-frequency acoustic collective oscillations occurs \cite{Platzman1973,Abstreiter1984}, as is illustrated in panels (a) and (c) of Fig. \ref{PI_ri_wq_25K} for $\mathcal{R}e~\Pi_{\rm e}(q,\omega)$ and $\mathcal{I}m~\Pi_{\rm e}(q,\omega)$, and in (b) and (d) for $\mathcal{R}e~\Pi_{\rm hh}(q,\omega)$ and $\mathcal{I}m ~\Pi_{\rm hh}(q,\omega)$.
The mode of acoustic collective oscillation is expected to be Landau damped due to single-particle excitations of the lighter mass species (conduction electrons), since the acoustic branch is lying within the single-particle excitation regime of the lighter species of the mcp as clearly demonstrated in Fig.\ref{epsilon}. 
The well defined optic modes dominated by the lighter species are intact to be clearly seen in $\mathcal{R}e~\Pi_{e}(q,\omega)$ and $\mathcal{I}m~\Pi_{e}(q,\omega)$ occurring at frequencies higher than that of the bare LO phonons well outside the single-particle excitation continuum.
The longitudinal low-frequency (optic) bare modes of the heavier species are drastically screened by the lighter mass species to become acoustic \cite{Abstreiter1984}.
The low-frequency acoustic plasmon branch is located outside single-particle excitations of heavy holes but is more broadened compared to the optic one. 
The contribution of the heavier mass species to dressed $\Pi_{hh} (q,\omega)$ [panels (b) and (d) of Fig. \ref{PI_ri_wq_25K}] is very similar to that of bare polarization functions $\Pi_c^0 (q,\omega)$ given by Eq.(\ref{Pi0}) and shown in Fig. \ref{P0_ri_wq_25K}.

\begin{figure*}[t] 
\includegraphics*[width=.9\textwidth]{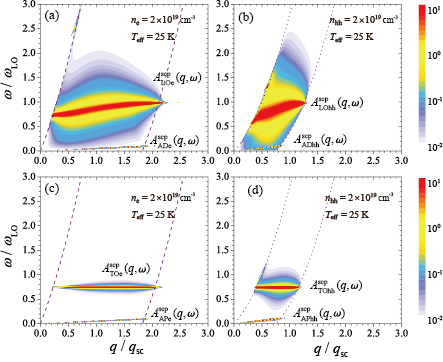}
\caption{(Color Online) Phonon spectral function $\mathcal{A}_{j\nu}^{\rm scp}(q,\omega)$ of a single-component plasma formed by  conduction electrons [(a) and (c)] and heavy holes [(b) and (d)] of concentration $2 \times 10^{19} \rm cm^{-3}$ at effective carrier temperature 25 K.
(a) and (b): Contributions from polar optical (LO) phonon and nonpolar acoustic deformation (AD) potential.
(c) and (d): Contributions from nonpolar optical (TO) and polar piezoacoustic (AP) deformation potentials.}
\label{A_e_scp}
\end{figure*}

The peak structure of the phonon spectral function $\mathcal{A}_{j\nu}(q,\omega)$ in the $\omega$-$q$ plane signifies the spectral behavior of renormalized phonon--plasmon coupled branches with collisional broadening.
In the presence of plasmon--phonon coupling, multiple peak structure is expected in the phonon spectral function of the material. 
The mode coupling of LO phonons and plasmons introduces a pair of branches named $L^{(+)}(\omega,q)$ and $L^{(-)}(\omega,q)$, the former (latter) representing high-frequency (low-frequency) coupled mode.
Detailed analyses of the longitudinal coupled modes have been reported extensively in the past for a single-component solid state plasma \cite{Cochran,Cohen}. 
In Fig. \ref{A_e_scp}, the spectral behaviors of the phonon spectral functions $\mathcal{A}_{j\nu}^{\rm scp}(q,\omega)$ in an electron plasma are illustrated for carrier concentration $2 \times 10^{19} \rm cm^{-3}$ at effective temperature 25 K.
The contributions of conduction electrons or of heavy holes to LO phonon spectral functions [$\mathcal{A}_{\rm LOe}^{\rm scp}(q,\omega)$ and $\mathcal{A}_{\rm LOhh}^{\rm scp}(q,\omega)$] and to acoustic phonon spectral functions [$\mathcal{A}_{\rm ADe}^{\rm scp}(\omega)$ and $\mathcal{A}_{\rm ADhh}^{\rm scp}(\omega)$] are displayed in panels (a) and (b).
Contributions to those of TO phonons [$\mathcal{A}_{\rm TOe}^{\rm scp}(q,\omega)$ and $\mathcal{A}_{\rm TOhh}^{\rm scp}(q,\omega)$] and of piezoacoustic deformation potentials [$\mathcal{A}_{\rm APe}^{\rm scp}(\omega)$ and $\mathcal{A}_{\rm APhh}^{\rm scp}(\omega)$] are illustrated in (c) and (d).
We note that latter phonon modes are negligibly sensitive to the presence of longitudinal plasmons resulting in narrow peaks near the corresponding bare phonon frequencies $\omega_{\rm TO}$ and $\omega_{\rm AC} (\sim sq)$ due to the absence of density fluctuations accompanied by the macroscopic electric fields associated with these nonpolar TO and acoustic modes. 
For the case of highly doped light-mass conduction electron plasma (of $n_{\rm e} =2 \times 10^{19} \rm cm^{-3}$ and $\omega_{\rm p,e}\gg \omega_{\rm LO}$) coupled with LO phonons [panel (a)], the $L^{(+)}(\omega,q)$ branch of $\tilde{\omega}_+(q) (\ge \omega_{\rm p,e})$ is highly plasmon-like with pronounced dispersion, but the $L^{(-)}(\omega,q)$ mode $\tilde{\omega}_-(q)$ shows strong phonon-like behavior with broad peaks for frequencies between $\omega_{\rm LO} > \omega >\omega_{\rm TO}$.   
Near $q=0$, the coupled phonon--electron plasma branches begin with frequencies $\tilde{\omega}_-\approx~0.7\omega_{\rm LO}$ and $\tilde{\omega}_+\approx~1.7\omega_{\rm LO}$.
We observe that, when $\omega_{\rm p,e} \gg \omega_{\rm LO}$, long-ranged Coulomb part of the LO mode is completely screened, in the long wave length limit, by the light-mass species (conduction electrons) becoming $\tilde{\omega}_-(q) \rightarrow \omega_{\rm TO} (\simeq 0.72\omega_{\rm LO})$.
The maximum in $\mathcal{A}_{j\nu}^{\rm scp}(q,\omega)$ of the phonon-like coupled branch of $\tilde{\omega}_-(q)$ appears at frequencies between $\omega_{\rm TO}$ and $\omega_{\rm LO}$ approaching asymptotically to $\omega_{\rm LO}$ subject to Landau damping as the wave number is increased \cite{Abstreiter1984}.
The higher frequency plasmon-like coupled branch of $\tilde{\omega}_+(q)$ starts at $q=0$ with $\omega_{\rm p,e} (\sim 1.7\omega_{\rm LO})$ showing strong dispersive behavior of relatively weaker strength.  
\begin{figure*} [t]
\includegraphics*[width=.9\textwidth]{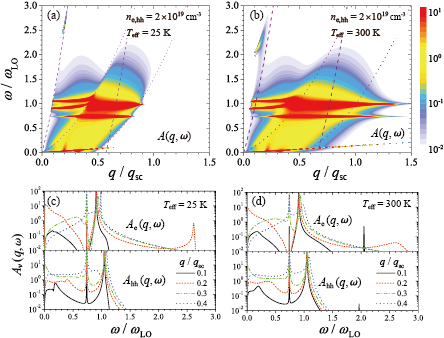}
\caption{(Color Online) Effective phonon spectral function $\mathcal{A}(q,\omega)$ of a two-component plasma with conduction electron concentration of $2 \times 10^{19} \rm cm^{-3}$.
(a) and (b): Dressed phonon spectral function at effective carrier temperatures 25 K and 300 K.
(c) and (d): Cross-sectional view of species-resolved phonon spectral function $\mathcal{A}_{\nu}(q,\omega)$ ($\nu=\rm e, hh$) at four different values of wave number $q$.
Contributions via electron-phonon coupling channels of polar (LO) and nonpolar (TO) optical phonons, and of piezoacoustic (AP) and nonpolar acoustic deformation (AD) potentials are summed.
}
\label{A_sum}
\end{figure*}

On the other hand, for the case of heavy-mass hole plasma (of $n_{\rm hh} =2 \times 10^{19} \rm cm^{-3}$ and $\omega_{\rm p,hh}\ll \omega_{\rm LO}$) coupled with LO phonons [panel (b)], the $L^{(+)}(\omega,q)$ branch is highly phonon-like starting $\tilde{\omega}_+(q\rightarrow 0)\sim\omega_{\rm LO}$, 
while the plasmon-like $L^{(-)}(\omega,q)$ coupled branch starts at $q=0$ with $\tilde{\omega}_-(0) \sim 0.4~\omega_{\rm LO}~(\ll \omega_{\rm TO})$ and approaches asymptotically to $\omega_{\rm LO}$. 
The latter heavy-mass species--phonon coupled mode shows larger broadening compared to the case of light-mass (electron) plasma [panel (a)].
The maximum in $\mathcal{A}_{j\nu}^{\rm scp}(q,\omega)$ of the phonon-like $L^{(+)}(\omega,q)$ coupled branch shows non-dispersive behavior of frequencies $\sim \omega_{\rm LO}$ with very minor strength. 
The behavior of $L^{(+)}(\omega,q)$ branch reflects the fact that, since $\omega_{\rm p,hh} < \omega_{\rm LO}$, the heavier mass hole plasma of low bare frequency is not fast enough to screen effectively the long-ranged Coulombic part of the LO oscillation, while the phonon gas is fast enough to screen the hole plasma depressing the plasmon oscillations \cite{Mahan72}.
The $L^{(-)}(\omega,q)$ plasmonic branch is located in the region of quasi-static screening due to $\omega_{\rm p,hh}\ll \omega_{\rm LO}$. 
Within the continuum region of single-particle excitations, modes are broadened and ill-defined because they are subject to Landau damping. 
For large wave numbers outside the single-particle excitation continuum, the screening is ineffective and, hence, effectively bare modes are resumed with sharp peaks.
The frequency of the coupled LO modes tends to $\omega_{\rm LO}$ for $q \gg q_{\rm sc}$. (See panels (a) and (b) of Fig. \ref{A_e_scp}.)

\begin{figure*} [t] 
\includegraphics*[width=.9\textwidth]{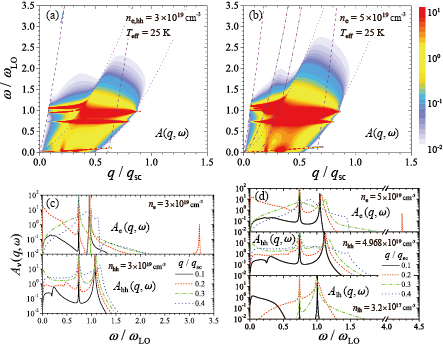}
\caption{(Color Online) Spectral behaviors of effective phonon spectral functions $\mathcal{A}(q,\omega)$ of a two-component plasma (2cp) having conduction electron concentration of $3 \times 10^{19} \rm cm^{-3}$ and of a three-component plasma (3cp) having conduction electron concentration of $5 \times 10^{19} \rm cm^{-3}$ at effective carrier temperature 25 K.
(a) and (b): Phonon spectral functions of $\mathcal{A}(q,\omega)$.
(c) and (d): Cross-sectional view of species-resolved $\mathcal{A}_{\nu}^{\rm 2cp}(q,\omega)$ ($\nu= \rm e,~hh$) and $\mathcal{A}_{\nu}^{\rm 3cp}(q,\omega)$  ($\nu= \rm e,~hh,~lh$) at three different values of $q$.
Contributions via electron--phonon coupling channels of polar (LO) and nonpolar (TO) optical phonons, and of piezoacoustic (AP) and nonpolar acoustic deformation (AD) potentials are added.
}
\label{3cp}
\end{figure*}

In a multi-component solid state plasma, the phonon spectral function would reveal multi-component features of the plasma -- multiple sets of phonon--plasmon coupled modes -- and their spectral behavior would become so involved to be resolved in the same way as the case of the scp.
For optically generated plasmas of weak excitation having not too much carrier concentration, for example, $\sim 10^{19} \rm cm^{-3}$, one can have a 2cp consisting of conduction electrons and heavy holes.
In Fig. \ref{A_sum} the effective phonon spectral function $\mathcal{A}(q,\omega)$ of a 2cp is shown for conduction electron concentration of $2 \times 10^{19} \rm cm^{-3}$ at effective carrier temperatures 25 K and 300 K, respectively.
Species-resolved frequency dependences are displayed separately in panels (c) and (d) at 25 K and 300 K for the purpose of spectral analysis.
The cross-sectional view of $\mathcal{A}_{\rm e}(q,\omega)$ and $\mathcal{A}_{\rm hh}(q,\omega)$ are illustrated for four different values of $q$.
Two sets of LO phonon--plasmon coupled modes ($\tilde{\omega}_{\pm}$) are seen in addition to the TO modes of $\omega_{\rm TO} \simeq 0.72 \omega_{\rm LO}$.
Boundaries of allowed single-particle excitations for electrons (holes) are designated by a pair of dashed (dotted) lines in each figure.
At $n_{\rm e} =2 \times 10^{19} \rm cm^{-3}$, the bare plasmon frequencies are $\omega_{\rm p,e}(\epsilon_\infty) =  1.68\, \omega_{\rm LO}$ and $\omega_{\rm p,hh}(\epsilon_\infty) = 0.69 \,\omega_{\rm LO}$ for the electrons and heavy holes, respectively, well separated from both optical phonon modes $L^{(+)}$ and $L^{(-)}$.
In a 2cp, there occur both optic and acoustic plasmons, and thus one can expect not only the individual LO phonon--optic plasmon coupled spectral behaviors of lighter and heavier species as illustrated in panels (a) and (b) of Fig. \ref{A_e_scp}, but also the contribution from the acoustic plasmon modes \cite{Pines,Abstreiter1984}. 
Our result shows that the multiple plasma species give substantial influence on the spectral behavior of the phonon spectral function.
Couplings of conduction electrons and heavy holes to both optical and acoustical phonon spectral functions appear simultaneously in the case of 2cp.
In panels (a) and (b), contributions to the spectral functions $\mathcal{A}_{\rm AD\nu}(q,\omega)$ of the deformation-potential induced acoustic phonon and $\mathcal{A}_{\rm AP\nu}(q,\omega)$ of piezoacoustic Coulomb interaction are also illustrated showing sharp peaks around the low-frequency bare $\omega_{\rm AC} (q)$. 
For the case of piezoelectricity induced Coulomb interaction, $A_{\rm AP\nu}(q,\omega)$ shows little but slightly enhanced collisional broadening compared to $\mathcal{A}_{\rm AD\nu}(q,\omega)$ of acoustic deformation-potential mechanism. 
The spectral intensity for the high-frequency plasmon-like $\tilde{\omega}_+^{\rm (e)} (q)$ conduction electron--phonon couples modes shows strong dispersive behavior starting with $\omega_{\rm p,e} (\sim 1.7\omega_{\rm LO})$ at $q=0$ and approaching the boundary of electronic single-particle excitation continuum. 
The spectral intensities for phonon-like coupled branches of $\tilde{\omega}_-^{(\rm e)} (q)$ and $\tilde{\omega}_+^{(\rm hh)} (q)$ are of very weak dispersion with peaks at frequencies close to $\omega_{\rm LO}$ and $\omega_{\rm TO}$.
Both branches show very sharp peak structures and vanish for the wave numbers beyond $q\simeq 1.5\,q_{\rm sc}$.
For the low-frequency plasmon-like $\tilde{\omega}_-^{(\rm hh)} (q)$ phonon--heavy hole plasma coupled mode, we understand that the spectral intensity of the mode is strongly suppressed because the low-frequency heavy-mass species plasma oscillation is so heavily screened by the light-mass species (conduction electrons) that the self-sustaining oscillations of hole plasmons are not permitted any more in this quasi-static region \cite{quasi-static}. 
We conjecture that additional little spectral strength with some broadening near the zero frequency peaked at around $q/q_{\rm sc}\sim 0.2$ is the so-called QPE-like branch \cite{DasSarma1990}, which appears in the region of finite $\mathcal{I}m~\Pi_{\nu\nu}$ and $\mathcal{R}e~\Pi_{\nu\nu} <0$ inside the continuum of single-particle excitations.
We observe that individual dressed branches are well resolved with slight overlap as seen in the panels (c) and (d).     
            
\begin{figure*}[t]  
\includegraphics*[width=.9\textwidth]{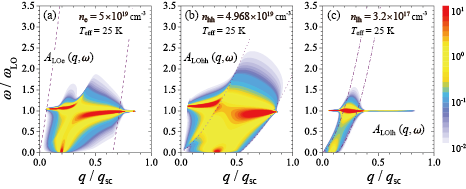}
\caption{(Color Online) Species-resolved phonon spectral function $\mathcal{A}_{\rm LO\nu}(q,\omega)$ ($\nu= \rm e,~hh,~lh$) of a three-component plasma formed by conduction electrons of concentration $5 \times 10^{19} \rm cm^{-3}$, heavy holes of concentration $4.968 \times 10^{19} \rm cm^{-3}$, and light holes of concentration $3.2 \times 10^{17} \rm cm^{-3}$ at effective carrier temperature 25 K.
(a) Contributions of conduction electron plasmon-LO phonon coupling.
(b) Contributions of heavy hole plasmon-LO phonon coupling.
(c) Contributions of light hole plasmon-LO phonon coupling.}
\label{n5-A_LO}
\includegraphics*[width=.9\textwidth]{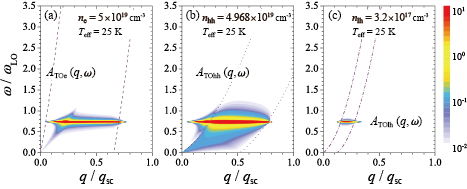}
\caption{(Color Online) Species-resolved phonon spectral function $\mathcal{A}_{\rm TO\nu}(q,\omega)$ ($\nu= \rm e,~hh,~lh$) of a three-component plasma formed by conduction electrons of concentration $5 \times 10^{19} \rm cm^{-3}$, heavy holes of concentration $4.968 \times 10^{19} \rm cm^{-3}$, and light holes of concentration $3.2 \times 10^{17} \rm cm^{-3}$  at effective carrier temperature 25 K.
(a) Contributions of conduction electron plasmon--TO phonon coupling.
(b) Contributions of heavy hole plasmon--TO phonon coupling.
(c) Contributions of light hole plasmon--TO phonon coupling.
}
\label{n5-A_TO}
\end{figure*}       

In Fig. \ref{3cp}, spectral behaviors of effective phonon spectral functions $\mathcal{A}(q,\omega)$ are compared for the cases of 2cp having conduction electron concentration $3 \times 10^{19} \rm cm^{-3}$ [panels (a) and (c)] and of three-component plasma (3cp) having conduction electron concentration $5 \times 10^{19} \rm cm^{-3}$ [panels (b) and (d)] at effective carrier temperature 25 K.
The contributions via electron--phonon coupling channels of polar LO and nonpolar TO phonons, and of piezoacoustic and nonpolar acoustic deformation potentials are summed.
Cross-sectional views of $\mathcal{A}_{\nu}^{\rm 2cp}(q,\omega)$ ($\nu=\rm e,~ hh$) for $3 \times 10^{19} \rm cm^{-3}$ and $\mathcal{A}_{\nu}^{\rm 3cp}(q,\omega)$ ($\nu=\rm e, ~hh, ~lh$) for $5 \times 10^{19} \rm cm^{-3}$ are illustrated in panels (c) and (d), respectively, at four different values of wave number $q$.
In a mcp of increased carrier concentrations, the compressive (optical) plasmon-like coupled modes begin with higher frequencies (for example, $\omega \ge 3 \omega_{\rm LO}$ for $3 \times 10^{19} \rm cm^{-3}$) at $q=0$, and the corresponding branches of very weak strengths are seen at $\omega \sim 3.2 \omega_{\rm LO}$ for $3 \times 10^{19} \rm cm^{-3}$ [panels (a) and (c)] and $\omega \sim 4.2 \omega_{\rm LO}$ for $5 \times 10^{19} \rm cm^{-3}$ [panel (d)] at $q\simeq 0.2q_{\rm sc}$. 
At 25 K, $q_{\rm sc} = 3.15\times 10^7 \rm cm^{-1}$ ($3.49\times 10^7 \rm cm^{-1}$) for an electron-hole plasma of conduction electron concentrations $3 \times 10^{19} \rm cm^{-3}$ ($5 \times 10^{19} \rm cm^{-3}$).          
The spectral behavior shown in panels (a) and (c) for $3 \times 10^{19} \rm cm^{-3}$ is very similar to a weakly excited case of  $2 \times 10^{19} \rm cm^{-3}$ shown in Fig. \ref{A_sum}(a) and (c), except the fact that the branch $\tilde{\omega}_+^{\rm (e)} (q)$ occurs at much increased frequency, for example, $\omega \sim 3.2~\omega_{\rm LO}$ at $q\simeq 0.2 q_{\rm sc}$.
For a photo-generated 3cp having conduction electrons of $5 \times 10^{19} \rm cm^{-3}$ at effective carrier temperature 25 K, the charge carriers of the plasma can be resolved into conduction electrons of concentration $5 \times 10^{19} \rm cm^{-3}$, heavy holes of concentration $4.968 \times 10^{19} \rm cm^{-3}$, and light holes of concentration $3.2 \times 10^{17} \rm cm^{-3}$.
Species-resolved phonon spectral function $\mathcal{A}_{\rm LO\nu}(q,\omega)$ and $\mathcal{A}_{\rm TO\nu}(q,\omega)$ of a 3cp ($\nu$=e, hh, and lh) with conduction electron concentration $5 \times 10^{19} \rm cm^{-3}$ are displayed in Figs. \ref{n5-A_LO} and \ref{n5-A_TO}.  
Boundaries of each individual allowed single-particle excitations are designated by a pair of dashed lines in each panel.
It is seen that, with the present choice of sample parameters, the effects of collisional broadening through carrier screening are not significant on the nonpolar phonon spectral functions $\mathcal{A}_{\rm TO\nu}(q,\omega)$ ($\nu=\rm e,~ hh,~ lh$) and $\mathcal{A}_{\rm LOlh}(q,\omega)$ because of very low concentration of light hole carriers.  
\begin{figure*} 
\includegraphics*[width=.9\textwidth]{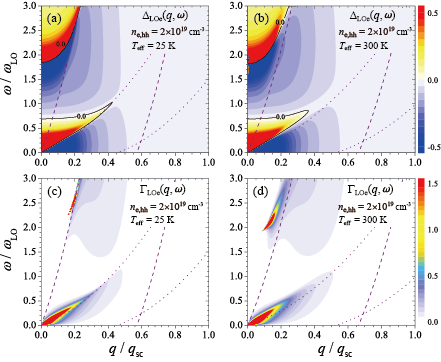}
\caption{(Color Online) Phonon self-energy correction $\mathcal{P}_{\rm LOe}(q,\omega)(\equiv \Delta_{\rm LOe}-i\Gamma_{\rm LOe}/2)$ in a two-component plasma via LO phonon--conduction electron coupling for concentration of $2 \times 10^{19} \rm cm^{-3}$ for each species.
(a) and (b): Real part of the self-energy correction $\Delta_{\rm LOe}(q,\omega)$ at 25 K and 300 K, respectively.
(c) and (d): Imaginary parts of the self-energy correction $\Gamma(q,\omega)$ at 25 K and 300 K, respectively. }
\label{self-energy}
\end{figure*}
\begin{figure*} [t] 
\includegraphics*[width=.9\textwidth]{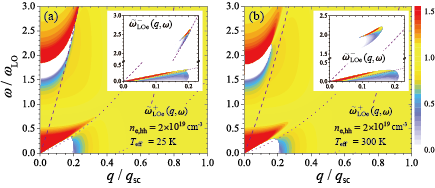}
\caption{(Color Online) Spectral behavior of the plasmon--phonon coupled modes $\tilde{\omega}_{\rm LOe}^{\pm}(q,\omega)$ in a photo-generated two-component plasma of conduction electron concentrations $2 \times 10^{19} \rm cm^{-3}$ at (a) 25 K and (b) 300 K.}
\label{dressed modes}
\includegraphics*[width=.9\textwidth]{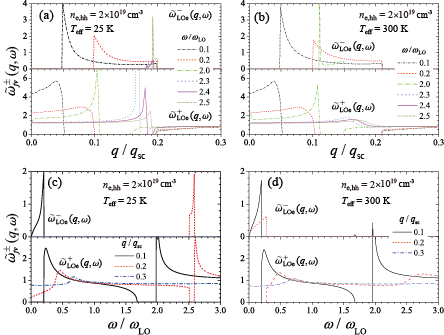}
\caption{(Color Online) Cross-sectional view of renormalized phonon modes ${\tilde\omega}_{\rm LOe}^+(q,\omega)$ and ${\tilde\omega}_{\rm LOe}^-(q,\omega)$ for electron concentrations of $2 \times 10^{19} \rm cm^{-3}$
(a) and (b): Behavior at representative values of frequency $\omega$ at 25 K and 300 K, respectively.
(c) and (d): Behavior at representative values of wave number $q$ at 25 K and 300 K, respectively.}
\label{SEq-SEw}
\end{figure*}

In Fig.\ref{self-energy}, the frequency and wave-number dependences of phonon self-energy correction $\mathcal{P}_{\rm LOe}(q,\omega)[\equiv\Delta_{\rm LOe}(q,\omega)-i\Gamma_{\rm LOe}(q,\omega)/2]$ are shown at 25 K and 300 K, respectively, for a photo-generated 2cp formed of conduction electrons and heavy holes each with concentration $2 \times 10^{19} \rm cm^{-3}$.
Contributions of the light-mass species (conduction electrons) $\Delta_{\rm LOe}(q,\omega)$ and $\Gamma_{\rm LOe}(q,\omega)$ are illustrated, respectively, in the $\omega$-$q$ space.
The contours of $\Delta_{\rm LOe}(q,\omega)=0$ are indicated by thin solid lines in panels (a) and (b). 
The spectral behaviors of $\mathcal{P}_{j\nu}(q,\omega)$ reveals very much similar structure as that of the dressed polarization function $\Pi_{\nu\nu}(q,\omega)$ of the plasma species, 
since $\Delta_{\rm j\nu}$ and $\Gamma_{\rm j\nu}$ are proportional to $\mathcal{R}e~\Pi_{\nu\nu}$ and $\mathcal{I}m~\Pi_{\nu\nu}$, respectively. [See Fig. \ref{PI_ri_wq_25K}(a) and (c).]
The sign of $\Delta_{j\nu}(q,\omega)$ is the same as that of ${R}e~\Pi_{\nu\nu}(q,\omega)$.
The real part of the self-energy correction $\Delta_{\rm LOe}(q,\omega)$ also changes signs crossing the zeros of $\mathcal{R}e~ \Pi_{\nu\nu}(q,\omega)$ with peaks just above and below the branches of high-frequency optic and low-frequency acoustic plasmon modes. 
On the other hand, $\Gamma_{j\nu}(q,\omega)$ is defined to be nonnegative and shows peak structure along the well-defined high-frequency compressive optic and low-frequency acoustic plasmon branches near the plasma cut-offs. 
We observe that the effect of plasma temperature on the phonon frequency renormalization [panels (a) and (b)] is moderate,
but the broadening in the self-energy correction [panels (c) and (d)] is more pronounced at higher temperature as was also seen in the case of the spectral function $\mathcal{A}(q,\omega)$ illustrated in Fig. \ref{A_sum}(a) and (b).  
The phonon self-energy corrections $\mathcal{P}_{\rm LOe}(q,\omega)$ of a photo-generated mcp for two different conduction electron concentrations $n_{\rm e}=3 \times 10^{19} \rm cm^{-3}$ and $5 \times 10^{19} \rm cm^{-3}$ at 25 K are given in Appendix B. (See Fig. \ref{n3-n5-self-energy}.)
In the case of conduction electron concentration $3 \times 10^{19} \rm cm^{-3}$ [panels (a) and (c) of Fig. \ref{n3-n5-self-energy}], the plasma consists of conduction electrons and heavy holes forming a 2cp and 
the spectral behavior of phonon self-energy correction $\mathcal{P}_{\rm LOe}(q,\omega)$ is very close to that of conduction electron concentration $2 \times 10^{19} \rm cm^{-3}$ shown in Fig. \ref{self-energy}.
Compelling difference is that the bare frequencies of both optic and acoustic plasmon modes are increased accordingly in a mcp as the carrier concentration is increased. 
For a plasma of conduction electron concentration $5 \times 10^{19} \rm cm^{-3}$, the light hole band is also occupied becoming a 3cp of $n_{\rm hh}=4.968 \times 10^{19} \rm cm^{-3}$ and $n_{\rm lh}=3.2 \times 10^{17} \rm cm^{-3}$.  
Panels (b) and (d) of Fig. \ref{n3-n5-self-energy} illustrate the spectral behavior of $\Delta_{\rm j\nu}$ and $\Gamma_{\rm j\nu}$, respectively, of a 3cp with conduction electron concentration $n_{\rm e}=5 \times 10^{19} \rm cm^{-3}$.
The effect of light hole carriers on $\mathcal{R}e~\Pi_{\rm ee}$ and $\mathcal{I}m~\Pi_{\rm ee}$ and, thus, to the LO phonon self-energy corrections $\Delta_{\rm LOe}$ and $\Gamma_{\rm LOe}$ are found to be negligible due to relatively too small concentration of light holes.

 In Fig. \ref{dressed modes}, spectral behaviors of the renormalized LO phonon frequencies ${\tilde\omega}_{\rm LOe}^+(q,\omega)$ and ${\tilde\omega}_{\rm LOe}^-(q,\omega)$ are shown for a photo-generated 2cp of electron concentration $2 \times 10^{19} \rm cm^{-3}$ at 25 K and 300 K, respectively.
The primary mode ${\tilde\omega}_{\rm LOe}^+(q,\omega)$ occurs in the domain of $2\Delta_{\rm LOe} > -\omega_{\rm LO}$, while the mode ${\tilde\omega}_{\rm LOe}^-(q,\omega)$ in the domain of $2\Delta_{\rm LOe} < -\omega_{\rm LO}$ only with finite values of $\Gamma_{\rm LOe}$ and negative $\Delta_{\rm LOe}$.
The latter mode is expected to be observable in the case of large $\Gamma_{\rm LOe}$.
In Fig. \ref{SEq-SEw}, renormalized phonon modes $\tilde\omega_{\rm LOe}^+(q,\omega)$ and $\tilde\omega_{\rm LOe}^-(q,\omega)$ shown in Fig. \ref{dressed modes} are resolved for representative values of frequency $\omega$ [panel (a) and (b)] and of wave number $q$ [panel (c) and (d)], respectively.
The frequency and wave number are scaled by the longitudinal bare phonon frequency $\omega_{\rm LO}$ and Thomas--Fermi screening wave number $q_{\rm sc}$, respectively.
We note that $\tilde\omega_{\rm LOe}^+(q,\omega)\simeq \omega_{\rm LO}$ over most of the domain satisfying the condition  $2 \Delta_{\rm LOe}(q,\omega) > -{\omega_{\rm LO}}$ in the $\omega$--$q$ plane except near the regions of well-defined plasmonic collective modes becoming $\tilde\omega_{\rm LOe}^+(q,\omega) >> \omega_{\rm LO}$.
Below the small opening gap [as indicated by white blank in panel (a)] of $2 \Delta_{\rm LOe}(q,\omega) < -{\omega_{\rm LO}}$, we observe the secondary mode $\tilde\omega_{\rm LOe}^-(q,\omega) < \omega_{\rm LO}$ over the domain designated in the insets. 
The mode $\tilde\omega_{\rm LOe}^-(q,\omega)$ is expected in the region of $2 \Delta_{\rm LOe}(q,\omega) < -{\omega_{\rm LO}}$ with finite values of $\Gamma_{\rm LOe}(q,\omega)$, as shown in Fig. \ref{self-energy}(c) and (d).                                             

Spectral behaviors of the renormalized phonon frequencies ${\tilde\omega}_{\rm LOe}^+(q,\omega)$ and ${\tilde\omega}_{\rm LOe}^-(q,\omega)$ are compared in Appendix B  for a 2cp of conduction electron concentrations $3 \times 10^{19} \rm cm^{-3}$ and a 3cp of $5 \times 10^{19} \rm cm^{-3}$, respectively.(See Fig. \ref{n3-n5-dressed modes}.)
The results illustrated in Fig. \ref{n3-n5-dressed modes}(a) and (b) are very similar to that of electron concentration $2 \times 10^{19} \rm cm^{-3}$ shown in Fig. \ref{dressed modes}(a) except the shifts in locations of the extrema along the frequency axis at a given effective carrier temperature. 
Cross-sectional views of the renormalized phonon modes ${\tilde\omega}_{\rm LOe}^+(q,\omega)$ and ${\tilde\omega}_{\rm LOe}^-(q,\omega)$ are given in Appendix B for representative values of frequency and wave number.(See Fig. \ref{n3-n5-SEq-SEw}.)
At increased conduction electron concentration of $5 \times 10^{19} \rm cm^{-3}$, the frequencies of collective modes of plasmon-phonon coupling shift to higher values, but the presence of light hole carriers of minor concentration is not seen to give observable effects in the spectral behavior of the phonon renormalization.

\section{Summary and Conclusion}
We have investigated spectral behavior of the dielectric response and phonon spectral functions of a mcp in an extended random phase approximation. 
The effects of dynamic screening, plasmon--phonon coupling, and exchange--correlations of the plasma species are included.
We have applied the formulation to the case of a photo-generated electron--hole plasma formed in an ideal wurtzite GaN, and scattering channels of various carrier-phonon couplings are considered.
Clear significance of the multiplicity of the plasma species is shown in the dielectric response and phonon spectral behaviors of a mcp. 

From the comparative study of the responses and phonon spectral functions of a mcp with that of a scp, we find following conclusions:
Dynamic screening and plasmon--phonon coupling are essential in understanding the spectral behavior of phonon spectral functions in a mcp. 
The effects of exchange and correlations among the carriers of the plasma are not seen significant for carrier concentrations $\sim 10^{19} \rm cm^{-3}$ shifting slightly down the frequencies of the optic and acoustic plasmon oscillations in the plasma.  

By extending linear response calculation to a mcp, we have detailed the spectral behavior of the dressed polarization functions  $\Pi_{\mu\nu}$ ($\nu$=e, hh) and examined plasma species-resolved dielectric functions $\epsilon_{\mu\nu}$.
A sum rule of $\sum_{\mu\nu} (\epsilon_{\mu\nu}^{-1}-\delta_{\mu\nu}) =0$ is found and multi-component character of the plasmonic oscillations was investigated.
From the zeros of the effective dielectric function $\mathcal{R}e ~\epsilon_{\rm eff}(q,\omega)$ of a 2cp, it is found that optic and acoustic plasmon branches are well separated from each other.
Hubbard-like local-field corrections of carriers are found to shift both branches slightly to reduced plasmon frequencies for given values of wave number. 
In a two-component electron--hole plasma, the higher frequency electron plasmon mode is almost intact and well defined near the bare plasma frequency $\omega_{\rm p,e}$. 
However, the lower frequency bare plasmon mode of heavier species (heavy holes) at $\omega_{\rm p,hh} $ is heavily screened by the lighter mass species (conduction electrons) resulting in an acoustic branch positioned inside the electron excitation continuum, the latter mode being subject to Landau damping by the light-mass species.
$\mathcal{I}m ~\epsilon_{\rm eff}(q,\omega)$ reveals double peak structure in a 2cp each peak appearing inside the single-particle excitation continua of electrons and holes of the plasma, unlike the case of a scp. 
The effects of local-field corrections of the carriers are appreciable only in the region of low frequency and long wavelength.

Dressed phonon propagators are evaluated in the $\omega$--$q$ plane and we have demonstrated that the dielectric screening in many carrier system gives rise to renormalized electron--phonon coupling modifying the phonon dispersion relations along with phonon spectral function.
The phonon frequency renormalization $\Delta_{j\nu}(q,\omega)$ and the phonon broadening $\Gamma_{j\nu}(q,\omega)$ are analyzed in terms of $\mathcal{R}e~\Pi_{\nu\nu}(q,\omega)$ and $\mathcal{I}m~\Pi_{\nu\nu}(q,\omega)$, respectively, and
it is found that the effect of plasma temperature on the phonon frequency renormalization is moderate, but the phonon broadening is more pronounced at higher temperature.
We have demonstrated that the plasmon--LO phonon coupling gives a pair of branches $L^{(+)}(\omega,q)$ and $L^{(-)}(\omega,q)$ and, in a 2cp, two sets of LO phonon--plasmon coupled modes ($\tilde{\omega}_{\pm}$) are seen. 
The spectral intensity for the high-frequency plasmon-like coupled mode $\tilde{\omega}_+^{\rm (e)} (q)$ shows strong dispersive behavior starting with $\omega_{\rm p,e}$ at $q=0$ and approaching the boundary of electronic single-particle excitation continuum. 
The spectral intensities for phonon-like coupled branches of $\tilde{\omega}_-^{(\rm e)} (q)$ and $\tilde{\omega}_+^{(\rm hh)} (q)$ are of very weak dispersion with peaks at frequencies close to $\omega_{\rm LO}$ and $\omega_{\rm TO}$.
The low-frequency plasmon-like coupled mode $\tilde{\omega}_-^{(\rm hh)}(q)$ is strongly suppressed so that the self-sustaining oscillations of hole plasmons are not permitted any more in this quasi-static region. 
Moreover, we have observed an additional peak of little spectral strength at $q\sim 0.2~q_{\rm sc}$ near the zero frequency in the domain of finite $\mathcal{I}m ~\Pi_{\nu\nu}$ and $\mathcal{R}e ~\Pi_{\nu\nu} <0$ inside the continuum of single-particle excitations.
The origin for the little peak is conjectured to be the so-called QPE-like branch \cite{DasSarma1990}. 
In a 3cp of conduction electron concentration $5 \times 10^{19} \rm cm^{-3}$,  
the effects of light hole carriers on $\mathcal{R}e~\Pi_{\rm ee}$ and $\mathcal{I}m~\Pi_{\rm ee}$ and, thus, to the LO phonon self-energy corrections $\Delta_{\rm LOe}$ and $\Gamma_{\rm LOe}$ are found to be negligible due to relatively too small concentration of light holes.
The spectral behaviors of the phonon self-energy correction reveals very much similar structure as that of the dressed polarization function $\Pi_{\nu\nu}(q,\omega)$ of the plasma species.
 
In conclusion, in a multi-component solid state plasma, multiple plasma species give substantial influences on the spectral behavior of the phonon spectral function, which is very distinct from that in a scp commonly seen in doped semiconductors.
The results of our computations are suitable for experimental observation, and we hope that the spectral behaviors demonstrated in the present work would be confirmed with various energy loss scattering measurements and hot carrier spectroscopies on photo-generated electron--hole plasmas in polar semiconductors.
Meaningful informations on the spectral behavior of a mcp with thermal and collisional broadening effects would be obtained by comparing experimental spectra with the results presented in the present paper.
The acoustic plasmon-like coupled modes of low frequency at long wavelength may be tested by ultrasonic measurements.  
 
\begin{acknowledgments}
The authors acknowledge the support in part by Basic Science Research Program (NRF-2013R1A1A4A01004433) through the NRF of Korea. One of the authors (HJK) acknowledges support by Priority Research Centers Program (2009-0093818) through the NRF of Korea.
We are grateful for stimulating discussions with E.H. Hwang on many-body correlations in a solid.
\end{acknowledgments}
          
\appendix
\begin{figure*} [t] 
\includegraphics*[width=.9\textwidth]{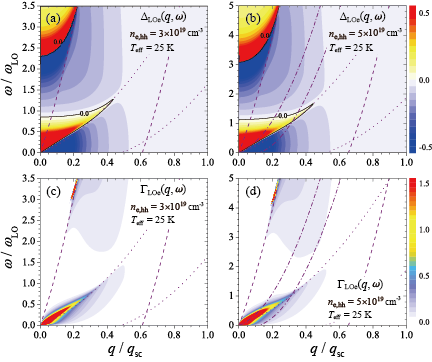}
\caption{(Color Online) Phonon self-energy corrections $\mathcal{P}_{\rm LOe}(q,\omega)(\equiv \Delta_{\rm LOe}-i\Gamma_{\rm LOe}/2)$ for a two-component plasma and a three-component plasma at 25 K.
(a) and (c): Real and imaginary parts of the self-energy correction for a two-component plasma with conduction electron concentration $3 \times 10^{19} \rm cm^{-3}$.
(b) and (d): Real and imaginary parts of the self-energy correction for a three-component plasma with conduction electron concentration $5 \times 10^{19} \rm cm^{-3}$.
}
\label{n3-n5-self-energy}
\end{figure*}
\section{Matrix elements of carrier-phonon coupling}
Here we briefly summarize matrix elements of carrier--phonon coupling channels employed in our discussion of spectral behavior of phonon spectral functions.
Of various channels of carrier--phonon scattering, we consider four distinct carrier-phonon coupling channels would-be important in compound semiconductors such as wurtzite GaN material  \cite{Conwell}:
couplings to polar longitudinal optical (LO) and nonpolar transverse optical (TO) phonons and couplings through acoustic deformation potential and piezoelectricity.  

In polar crystals, the longitudinal modes due to Fr\"ohlich-type Coulombic interaction induces a long ranged electrical polarization field.
For the case of carrier--polar optical phonon coupling, the squared coupling matrix element is given by \cite{Mahan}  
\be
|M_{q,\nu}^{\mathrm{LO}}(q)|^2 = \frac{4\pi \hbar}{q^2} \sqrt{\frac{\hbar^3\omega_{\rm LO}^3}{2m_\nu}} \alpha,
\label{M-LO}
\ee
where $\alpha$ is the dimensionless Fr\"ohlich coupling constant $\alpha=\frac{e^2}{\hbar}\sqrt{m_{\nu}/(\hbar\omega_{\rm LO})}\left(\frac{1}{\epsilon_\infty}-\frac{1}{\epsilon_0}\right)$,
and $\epsilon_\infty$ and $\epsilon_0$ are, respectively, the optical and static dielectric constants of the material.
In GaN, much enhanced carrier--polar phonon interaction is expected due to its higher ionicity giving, for example, the Fr\"ohlich coupling constant $\alpha (\rm GaN) \sim 6 \alpha (\rm GaAs)$ with  $\omega_{\rm LO} (\rm GaN) \sim 3 \omega_{\rm LO}(\rm GaAs)$ \cite{Mora1999}.
For crystals lacking a center of symmetry, acoustic phonons also induce
an electric polarization field giving rise to piezoelectricity through Coulombic interaction.
The piezoelectricity is most commonly found in the wurtzite structure, and the matrix element $M_{q,\nu}^{\mathrm{AP}}(q)$ for piezoelectric acoustic (AP) phonon coupling with frequency $\omega(q)$ and velocity $s(q)$ is given by \cite{Mahan72}
\begin{equation}
\mid M_{q,\nu}^{\mathrm{AP}}(q)\mid^2 = \frac{2\hbar s(q)}{\rho_{\nu} \omega(q) q^2 \epsilon^2 (q)}
\biggl[\sum_{ijk} q_k e_{k,ij} \xi_i(\lambda,q ) q_j \biggr]^2.
\label{M-AP} 
\end{equation}
Here $\rho_{\nu}$, $\epsilon(q)$, $\xi(\lambda,q)$, and $e_{k,ij}$ denote the
average mass density, dielectric constant of the material, the unit vector of the acoustic lattice
polarization $\lambda$ (= TA or LA), and the piezoelectric coupling constants of a third rank tensor, respectively \cite{Mahan72}.  
Because strain modifies local band structure of the material, the deformation potentials can also induce carrier--nonpolar TO and acoustical phonon couplings. 
The matrix element $M_{q,\nu}^{\mathrm{TO}}(q)$ for nonpolar optical phonon coupling is given, in terms of optical deformation potential, by \cite{Conwell}
\begin{equation}
\mid M_{q,\nu}^{\mathrm{TO}}(q)\mid^2 = \frac{\hbar \mathcal{D}^2}{ 2 \rho_{\nu} \omega_{\mathrm{TO}}\mathcal{V}}.
\label{M-TO}
\end{equation}
Here $\mathcal{D}$ and $\mathcal{V}$ are the optical deformation potential constant and the volume of the sample, respectively.
On the other hand, the matrix element $M_{q,\nu}^{\mathrm{AD}}(q)$ for acoustical deformation potential scattering is written as \cite{Conwell,Kittel}
\be
\mid M_{q,\nu}^{\mathrm{AD}}(q)\mid^2 = \frac{\hbar \mathcal{E}_\nu^2 q^2}{ 2 \rho_{\nu} \omega_q\mathcal{V}},.\label{M-ADP}
\ee
Here $\mathcal{E}_\nu$ is the deformation potential constants of the longitudinal acoustical waves for the carriers in the conduction and valence bands.

\section{Spectral behavior of phonon self-energy corrections and renormalized phonon modes of two-component and three-component plasmas}
In a photo-generated electron--hole plasma, the number of plasma species can be modulated as a function of the carrier concentration $n_{\rm e}$ in the conduction band. 
For higher values of $n_{\rm e}$, both valence bands of heavy holes and light holes can be occupied giving rise to 3cp.
For a plasma of conduction electron concentration $5 \times 10^{19} \rm cm^{-3}$, the light hole band is also occupied becoming a 3cp of $n_{\rm hh}=4.968 \times 10^{19} \rm cm^{-3}$ and $n_{\rm lh}=3.2 \times 10^{17} \rm cm^{-3}$.   
Phonon self-energy corrections $\mathcal{P}_{\rm LOe}(q,\omega)$ of a photo-generated mcp for two different conduction electron concentrations $n_{\rm e}=3 \times 10^{19} \rm cm^{-3}$ and $5 \times 10^{19} \rm cm^{-3}$ at 25 K are given in Fig. \ref{n3-n5-self-energy}.
In the case of conduction electron concentration $3 \times 10^{19} \rm cm^{-3}$ [panels (a) and (c) of Fig. \ref{n3-n5-self-energy}], the plasma consists of conduction electrons and heavy holes forming a 2cp and 
the spectral behavior of phonon self-energy correction $\mathcal{P}_{\rm LOe}(q,\omega)$ is very close to that of conduction electron concentration $2 \times 10^{19} \rm cm^{-3}$ shown in Fig. \ref{self-energy}.
Noticeable difference is that the bare frequencies of both optic and acoustic plasmon modes are increased accordingly in a mcp as the carrier concentration is increased. 
Panels (b) and (d) of Fig. \ref{n3-n5-self-energy} illustrate the spectral behavior of $\Delta_{\rm j\nu}$ and $\Gamma_{\rm j\nu}$, respectively, of a 3cp with conduction electron concentration $n_{\rm e}=5 \times 10^{19} \rm cm^{-3}$.
The effect of light hole carriers on $\mathcal{R}e~\Pi_{\rm ee}$ and $\mathcal{I}m~\Pi_{\rm ee}$ and, thus, to the LO phonon self-energy corrections $\Delta_{\rm LOe}$ and $\Gamma_{\rm LOe}$ are found to be negligible due to relatively too small concentration of light holes.

\begin{figure*} [h]  
\includegraphics*[width=.9\textwidth]{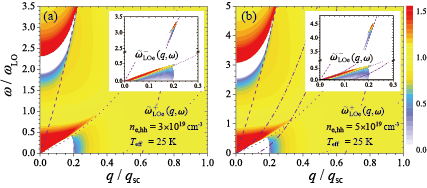}
\caption{(Color Online) Spectral behavior of the renormalized phonon frequencies  ${\tilde\omega}_{\rm LOe}^{\pm}(q,\omega)$ in a photo-generated plasma at 25 K of conduction electron concentrations (a) $3 \times 10^{19} \rm cm^{-3}$ and (b) $5 \times 10^{19} \rm cm^{-3}$.
}
\label{n3-n5-dressed modes}
\end{figure*}

Spectral behaviors of renormalized phonon frequencies ${\tilde\omega}_{\rm LOe}^+(q,\omega)$ and ${\tilde\omega}_{\rm LOe}^-(q,\omega)$ are compared in Fig. \ref{n3-n5-dressed modes} for a 2cp of conduction electron concentrations $3 \times 10^{19} \rm cm^{-3}$ and a 3cp of $5 \times 10^{19} \rm cm^{-3}$, respectively.
In panel (b), boundaries of the continuum of single-particle excitations for light holes are indicated by a pair of dash-dotted lines.
\begin{figure*} 
\includegraphics*[width=.9\textwidth]{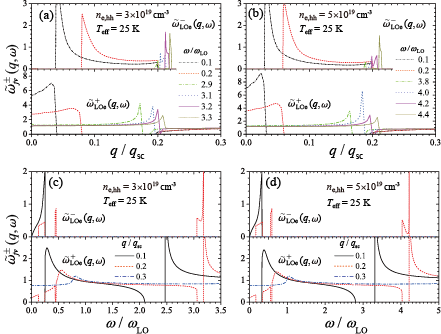}
\caption{(Color Online) Wave-number and frequency dependences of renormalized phonon modes ${\tilde\omega}_{\rm LOe}^{\pm}(q,\omega)$ at 25 K.
(a) and (b): Behavior for representative values of frequency $\omega$ for electron concentrations of $3 \times 10^{19} \rm cm^{-3}$ and $5 \times 10^{19} \rm cm^{-3}$, respectively.
(c) and (d): Behavior for representative values of wave number $q$ for electron concentrations of $3 \times 10^{19} \rm cm^{-3}$ and $5 \times 10^{19} \rm cm^{-3}$, respectively.
}
\label{n3-n5-SEq-SEw}
\end{figure*}
The frequency and wave number are scaled by the longitudinal bare phonon frequency $\omega_{\rm LO}$ and Thomas--Fermi screening wave number $q_{\rm sc}$, respectively.
For a photo-generated plasma with electron concentration $5 \times 10^{19} \rm cm^{-3}$, the presence of additional carriers of concentration $3.2 \times 10^{17} \rm cm^{-3}$ in the light hole band gives rise to the third species making the plasma a three-component one.
The results illustrated in Fig. \ref{n3-n5-dressed modes}(a) and (b) are very similar to that of electron concentration $2 \times 10^{19} \rm cm^{-3}$ shown in Fig. \ref{dressed modes}(a), the locations of the extrema are shifted along the frequency axis at a given effective carrier temperature. 
 In Fig. \ref{n3-n5-SEq-SEw}, cross-sectional views of the renormalized phonon modes ${\tilde\omega}_{\rm LOe}^+(q,\omega)$ and ${\tilde\omega}_{\rm LOe}^-(q,\omega)$ shown in Fig. \ref{n3-n5-dressed modes} are illustrated for representative values of frequency and wave number. 
In panels (a) and (b), the wave-number dependences of ${\tilde\omega}_{\rm LOe}^{\pm}(q,\omega)$ are shown for constant values of frequency $\omega$ at 25 K, while 
the frequency dependences at representative values of wavenumber $q$ are shown in panels (c) and (d) for electron concentrations of $3 \times 10^{19} \rm cm^{-3}$ and $5 \times 10^{19} \rm cm^{-3}$, respectively.
For conduction electron concentration of $5 \times 10^{19} \rm cm^{-3}$, the frequencies of plasmon--phonon coupled modes shift to higher values, but the presence of light hole carriers of minor concentration is not seen to give observable effects in the spectral behavior of the phonon renormalization.



\begin{thebibliography}{99}
\bibitem{Kyhm2011} K. Kyhm, L. Rota, and R.A. Taylor, Phys. Status Solidi A {\bf 208}, 1159 (2011).
\bibitem{Hagele} D. H\"agele, R. Zimmermann, M. Oestreich, M.R. Hofmann, and W.W. Ruhle, B.K. Meyer, H. Amano, and I. Akasaki, Phys. Rev. B {\bf 59}, 7797 (1999). 

\bibitem{Ozgur} U. Ozgur, Ya.I. Alivov, C. Liu, A. Teke, M.A. Reshchikov, S. Doan, V. Avrutin, S.-J. Cho, and H. Morkoc, J. Appl. Phys. {\bf 98}, 041301 (2005), and the references therein.

\bibitem{Sarua} A. Sarua, M. Kuball, and J.E. Van Nostrand, Appl. Phys. Lett.{\bf 85}, 2217 (2004).
\bibitem{Ishioka} K. Ishioka, Phys. Rev. B {\bf 84}, 235202 (2011). 
\bibitem{Quinn-Yi} J.J. Quinn and K.S. Yi, {\it Solid State Physics}, (Springer, Heidelberg, 2009), Ch. 8.
 
\bibitem{Abstreiter1984} G. Abstreiter, M. Cardona, and A. Pinczuk, {\it Light Scattering in Solids IV} ed. by M. Cardona and G. G\"untherodt, (Springer-Verlag, New York, 1984), p.5, and the references there in.
\bibitem{Pinczuk1977} For example, A. Pinczuk, G. Abstreiter, R. Trommer, and M. Cardona, Solid St. Commun.  {\bf 21}, 959 (1977), and the references therein.
\bibitem{Abstreiter1979} G. Abstreiter, R. Trommer, M. Cardona, and A. Pinczuk, Solid St. Commun.  {\bf 30}, 703 (1979), and the references therein.
\bibitem{Romanek} K.M. Romanek, H. Nather, and E.O. G\"obel,  Solid St. Commun. {\bf 39}, 23 (1981).
\bibitem{DasSarma1990} See, for example, S. Das Sarma, J.K. Jain, and R. Jalabert, Phys. Rev. B {\bf 41}, 3561 (1990).

\bibitem{Platzman1973} P.M. Platzman and P.A. Wolff, {\it Waves and Interactions in Solid State Plasmas} (Academic Press, New York, 1973).
\bibitem{Pinczuk1981} A. Pinczuk, J. Shah, and P.A. Wolff, Phys. Rev. Lett. {\bf 47}, 1487 (1981).
\bibitem{Esperidiao} A.S. Esperidiao, A.R. Vasconcelos, and R. Luzzi, Phys. Stat. Sol. (b) {\bf 168}, 533 (1991).
\bibitem{Pines} P. Nozi\'eres and D. Pines, {\it The Theory of Quantum Liquids}, (Perseus Books, Cambridge, MA, 1999).
\bibitem{Jain1988} J.K. Jain, R. Jalabert, and S. Das Sarma, Phys. Rev. Lett. {\bf 60}, 353 (1988).
\bibitem{Yi-Kim-CAP2015} K.S. Yi and H.-J. Kim, Curr. Appl. Phys. {\bf 15}, 335 (2015).
\bibitem{YI&Kim_PhyscaB}  K.S. Yi and H.J. Kim, Physica B {\bf 457}, 149 (2015).

\bibitem{Pau} S. Pau {\it et al.}, Phys. Rev. B {\bf 57}, 7066 (1998).
\bibitem{Landolt} Landolt-B\"ornstein: {\it Numerical Data and Functional Relationships in Science and Technology},
Vol. 17 Semiconductors, ed. by O. Madelung, M. Schulz and H. Weiss, (Springer Verlag, Berlin, 1982).
\bibitem{Grahn} H.T. Grahn, {\it Introduction to Semiconductor Physics} (World Scientific, Singapore, 1999).

\bibitem{Kyhm} K. Kyhm {\it et al.}, Phys. Rev. B {\bf 65}, 193102 (2002).
\bibitem{Gil} B. Gil, O. Briot, and R.L. Aulombard, Phys. Rev. B {\bf 52}, R17028 (1995).

\bibitem{Tchounkeu} M. Tchounkeu, O. Briot, B. Gil, J.P. Alexis, and R.L. Aulombard, J. Appl. Phys. {\bf 80}, 5352 (1996).

\bibitem{Chuang} S.L. Chuang and C.S. Chang, Phys. Rev. B {\bf 54}, 2491 (1996).

\bibitem{Walle} C.G. Van de Walle, Phys. Rev. B {\bf 68}, 165209 (2003).
\bibitem{Guess} Since optical deformation potential has not been known, we used $10^9$ eV/cm, which is usually known in III--V semiconductors.
\bibitem{Nakamura} S. Nakamura, S. Pearton, and G. Fasol, {\it The Blue Laser Diode: The Complete Story}, (Springer Verlag, Berlin, 2000).

\bibitem{Binet1999} F. Binet, J.Y. Duboz, J. Off, and F. Scholz, Phys. Rev. B {\bf 60}, 4715 (1999).
\bibitem{GaN}  Each band is assumed to have two-fold spin degeneracy with  $\Gamma^6_9$ (heavy), $\Gamma^6_7$ (light), and $\Gamma^1_7$ (split-off) symmetries, respectively.
\bibitem{Suzuki} M. Suzuki, T Uenoyama, and A. Yanase, Phys. Rev. B {\bf 52}, 8132 (1995).
\bibitem{Jeon} J.-B. Jeon, Y.M. Sirenko, K.W. Kim, M.A. Littlejohn, and M.A. Stroscio, Solid St. Commun. {\bf 99}, 423-426 (1996).

\bibitem{Bruus}  H. Bruus and K. Fensberg, {\it Many-Body Quantum Theory in Condensed Matter Physics}, (Oxford Univ. Press, New York, 2004).

\bibitem{ksyi2007} K.S. Yi, J.S. Kim, and K. Kyhm, J. Korean Phys. Soc. {\bf 50}, 1670 (2007).
\bibitem{Mahan72} G. Mahan, {\it Polarons in Ionic Crystals and Polar 
Semiconductors}, edited by J.T. Devreese (North Holland, Amsterdam, 1972), pp. 553.

\bibitem{Fetter} A. Fetter and J.D. Walecka, {\it Quantum Theory of Many-Particle Systems}, (McGraw-Hill, New York, 1971).
\bibitem{Kohn-Sham1965}  W. Kohn and L.J. Sham, Phys. Rev. {\bf 140}, A1133-8 (1965).
\bibitem{KukkonenOverhauser1979}  C.A. Kukkonen and A.W. Overhauser, Phys. Rev. B {\bf 20}, 550-557 (1979).
\bibitem{Hubbard} J. Hubbard, Proc. Roy. Soc. A {\bf 243}, 336 (1957); An example of Hubbard's local-field corrections is $G_{ij}(q )=\frac{1}{2} q^2/(q^2+k_{{\rm F}i}^2+q_{{\rm TF}i}^2)\delta_{ij}$, where $k_{{\rm F}i}$ and $q_{{\rm TF}i}$ are the Fermi and Thomas--Fermi wave numbers, respectively, of the $i^{\rm th}$ species of the multi-component plasma \cite{Mahan}. 
In this work, we have used this expression of $G_{ij}(q)$ extended to include ladder diagrams in our evaluation of $\psi_{ij}$ and $\Delta(q,\omega)$.
\bibitem{Vashishta1974}  P. Vashishta, P. Bhattacharyya, and K.S. Singwi, Phys. Rev. B {\bf 10}, 5108 (1974).
\bibitem{HedinLundqvist1971} L Hedin and B.I. Lundqvist, J. Phys. C: Solid St. Phys. {\bf 4}, 2064-2083 (1971).
\bibitem{Yi-Quinn1996} K.S. Yi and J.J. Quinn, Phys. Rev. B {\bf 54}, 13398 (1996).
\bibitem{Mahan} See, for example, G. Mahan, {\it Many Particle Physics}, 3rd ed. (Plenum, 
New York, 2000).
\bibitem{KimParkYi2011}  J.S. Kim, S. Park, and K.S. Yi, J. Korean Phys. Soc. {\bf 58}, 1429 (2011).
\bibitem{Maldague} P.F. Maldague, Surf. Science {\bf 73}, 296 (1978).
\bibitem{Giuliani} G. Giuliani and G. Vignale, {\it Quantum Theory of Electron Liquid}, (Cambridge Univ. Press, 
New York, 2005).

\bibitem{Collet1989} J.H. Collet, Phys. Rev. B {\bf 39}, 7659 (1989).
\bibitem{Hwang2007} E.H. Hwang and S. Das Sarma, Phys. Rev. B {\bf 75}, 205418 (2007).

\bibitem{ashcroft-mermin} N.W. Ashcroft, N.D. Mermin, {\it Solid State Physics}, (Brook-Cole Cengage Learning, New York, 1976).
\bibitem{Cochran} See, for example, W. Cochran, R.A. Cowley, G. Dolling, and M. M. Elcombe, Proc. Roy. Soc., Ser. A {\bf 293}, 433 (1966). The authors illustrated the dispersion relation of the coupled modes.
\bibitem{Cohen} See, for example, M. Cohen, in {\it Superconductivity}, edited by R. D. Parks (Dekker, New York, 1969), p.659. The author discussed the behavior of the coupled modes neglecting the damping effect in the long wavelength limit.
\bibitem{quasi-static} In this region, the dielectric screening is quasi-static, since the plasmon frequency of light-mass species is considerably greater than the frequency ${\tilde\omega}_-^{\rm (hh)} (q)$ of the plasmon-like heavy-mass species plasmon--LO phonon coupled mode.
\bibitem{Conwell} E. Conwell, {\it High Field Transport in Solids} (Academic, New York, 1963).
\bibitem{Mora1999} M.E. Mora-Ramos, F.J. Rodriguez, and J. Quiroga, J. Phys.: Condens. Matter {\bf 11}, 8223 (1999), and the references there in.
\bibitem{Kittel} C. Kittel, {\it Quantum Theory of Solids}  (Wiley, New York, 
1987).
\end{thebibliography}
\end{document}